\documentclass[english,a4paper,11pt]{article}
\usepackage[margin=3cm]{geometry}
\usepackage[latin1]{inputenc}
\usepackage{latexsym}
\usepackage{booktabs}
\usepackage{amsmath}
\usepackage{changepage}
\usepackage{placeins}
\usepackage{algorithm}
\usepackage{makecell}
\usepackage{algpseudocode}
\usepackage{latexsym}
\usepackage{pdflscape}
\usepackage[final]{graphicx}
\DeclareGraphicsExtensions{.jpg,.jpeg,.pdf,.png,.mps}
\usepackage{epsfig}
\usepackage[round]{natbib}
\usepackage{rotating}
\usepackage{color}
\usepackage{xcolor}
\usepackage{graphicx,subfig}
\usepackage{colortbl}
\usepackage{float}
\newcommand{\Tau}{\mathrm{T}}
\usepackage{hyperref}
\usepackage{csquotes}
\usepackage[misc]{ifsym}
\usepackage{multirow}
\usepackage[toc,page]{appendix}
 \usepackage{threeparttable}
\usepackage[bitstream-charter]{mathdesign}
\usepackage[T1]{fontenc}
\usepackage{lmodern}
\title{Short-term CO$_2$ emissions forecasting: insight from the Italian electricity market}

\author{$\mathrm{Pierdomenico \ Duttilo}^\mathrm{*,\hspace{0.5mm}\textrm{\Letter}},  
	\  \mathrm{Francesco \ Lisi}^\mathrm{*}$
	\\  
	$^\mathrm{*}$\small{\emph{Department  of Statistical Sciences,  University of Padua, Italy}}\\
    $^\mathrm{\textrm{\Letter}}$\small{\emph{Corresponding author: \href{mailto:pierdomenico.duttilo@unipd.it}{\textcolor{blue}{pierdomenico.duttilo@unipd.it}}}}
}
\date{}

\begin{document}
	\maketitle
\begin{abstract}
This study investigates the short-term forecasting of carbon emissions from electricity generation in the Italian power market. Using hourly data from 2021 to 2023, several statistical models and forecast combination methods are evaluated and compared at the national and zonal levels. Four main model classes are considered: (i) linear parametric models, such as seasonal autoregressive integrated moving average and its exogenous variable extension; (ii) functional parametric models, including seasonal functional autoregressive models, with and without exogenous variables; (iii) (semi) non-parametric and possibly non-linear models, notably the generalised additive model (GAM) and TBATS (trigonometric seasonality, Box-Cox transformation, ARMA errors, trend, and seasonality); and (iv) a semi-functional approach based on the K-nearest neighbours. Forecast combinations include simple averaging, the optimal Bates and Granger weighting scheme, and a selection-based strategy that chooses the best model for each hour. The results show that the GAM produces the most accurate forecasts during the daytime hours, while the functional parametric models perform best in the early morning. Among the combination methods, the simple average and the selection-based approaches consistently outperform all individual models. The findings underscore the value of hybrid forecasting frameworks in improving the accuracy and reliability of short-term carbon emissions predictions in power systems. In addition, they highlight the importance of considering zonal specificities when implementing flexible energy demand strategies, as the timing of low-carbon emissions varies between market zones throughout the day.
	
\noindent 
\hspace{1cm}
\end{abstract}

\section{Introduction}\label{sec.intro}
Carbon dioxide (CO$_2$) emissions are one of the main drivers of global climate change. According to the emissions database for global atmospheric research \citep{edgar2024}, in 2023, China, the world's largest emitter, accounted for 33.98\% of global CO$_2$ emissions, followed by the USA (12\%), India (7.57\%), Europe (6.44\%) and Russia (5.30\%). The power sector remains a key contributor, responsible for 38.24\% of total CO$_2$ emissions in the same year. In particular, the power sectors of China, the USA, India, Europe and Russia contributed 16.59\%, 3.75\%, 3.53\%, 2.26\%, and 1.64\% of global CO$_2$ emissions, respectively. As a result, these countries represented 72.61\% of global CO$_2$ emissions from the power sector.

The amount of CO$_2$ emissions from electricity generation depends on the share of renewable sources (e.g. hydro, solar, and wind) and nonrenewable sources (e.g. coal, oil, and natural gas). The mix of sources changes over time according to the mechanisms of the electricity market and weather conditions \citep{KOCAK2023}. Although renewable energy generation reached a global record of 927 TWh in 2024, CO$_2$ emissions from the power sector increased by 1.6\% compared to 2023. This increase was mainly driven by an increase in global electricity demand, particularly due to extreme heat in countries such as India \citep{carbonbrief2025}. The rise in fossil fuel use highlights the ongoing challenge of decarbonising under climate stress.

Over time, most countries have made efforts to reduce CO$_2$ emissions through international initiatives. The Paris Agreement, signed in 2015, aims to keep global warming well below $2^\circ$C above pre-industrial levels, with the aim of staying below $1.5^\circ$C \citep{UNFCCC2015}. It also sets specific targets for the power sector, including a 20\% reduction in CO$_2$ emissions, a 20\% increase in the use of renewable energy, and a 20\% improvement in energy efficiency worldwide. These actions also support Sustainable Development Goal 7 (SDG 7), introduced in the same year with the 2030 Agenda for Sustainable Development, which focusses on ensuring access to affordable, reliable, and clean energy for all \citep{SDG2015}. Making the power sector more sustainable is essential to reduce CO$_2$ emissions and address climate change.

As part of the global shift towards cleaner energy in line with the climate targets of the United Nations \citep{TARGET2050}, more attention is paid to strategies that reduce CO$_2$ emissions from electricity generation.\\
From the supply side, proposed solutions include the integration of renewable energy sources with energy storage systems to manage fluctuations in renewable electricity generation. Batteries, fuel cells, or hydro reservoirs can be programmed to charge during periods of high renewable generation, typically associated with low emissions, and discharge when renewable output is low and consequently emissions are higher \citep{HADDADIAN2015,IEA2019,He2024}.\\
On the demand side, there is growing interest in strategies that shift flexible energy demand from periods of high emissions to periods of low emissions. Smart charging represents an opportunity to reduce CO$_2$ emissions by scheduling charging sessions during periods of lower emissions in electricity generation \citep{JOCHEM2015}. For example, residential charging of electric vehicles can be strategically planned when CO$_2$ emissions are low \citep{huber2018,HUBER2021}. Even if electric vehicle drivers are not experts in the electric system \citep{BIRESSELIOGLU2018}, their willingness to adopt flexible charging, such as accepting delays, increases when it contributes to greater integration of renewable energy and reduces charging costs and carbon emissions \citep{WILL2016,jung2019}.

Consumers typically schedule electricity usage based on price forecasts to minimise costs \citep{voronin2013}. However, price-based scheduling can lead to lower emissions only when renewable energy production exceeds the required \citep{LEERBECK2020}. Otherwise, it may result in a phenomenon known in the literature as the \textit{``merit order emission dilemma''}, where low-cost but high-emission sources, such as coal, are preferred \citep{wagner2002merit,FLESCHUTZ2021}. Recently, more attention has been paid to the environmental impact of electricity consumption, with the aim of reducing CO$_2$ emissions by promoting a shift to lower-carbon energy sources. In this context, electricity usage could be scheduled not only based on predicted prices but also on predicted CO$_2$ emissions, as a primary or secondary objective \citep{BOKDE2021}. Therefore, accurate  CO$_2$ forecasts are essential to support emission-based scheduling of flexible demand and storage, at least until electricity generation becomes fully renewable. 

\begin{figure}[H]
    \centering
    \includegraphics[width=0.9\textwidth]{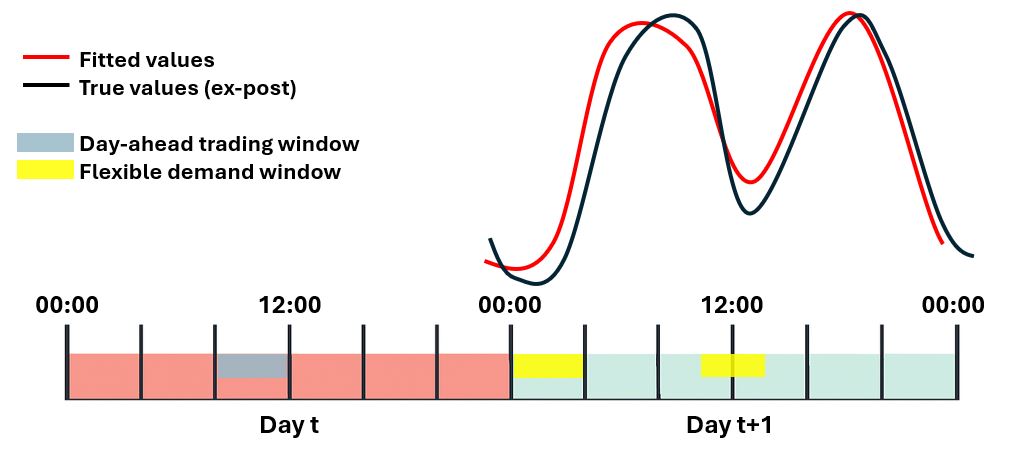}
    \caption{Time framework to forecast carbon emissions from electricity generation in the day-ahead market.}
    \label{fig:motivating_example}
\end{figure}

Figure~\ref{fig:motivating_example} shows the temporal structure of one day-ahead forecasts of carbon emissions from electricity generation in the day-ahead market. In this market, producers submit bids and offers for each hourly interval of day $t+1$. These transactions typically occur during the morning of day $t$, at which point the electricity generation for day $t+1$, as well as its associated carbon emissions, are unavailable. The red curve shows the fitted (forecasted) carbon emission profile for day $t+1$, while the black curve represents the true realisation observed ex-post. Accurate forecasts are crucial to identify low-carbon hours, namely periods during which electricity generation is expected to produce lower emissions. The grey highlighted intervals (00:00--06:00 and 11:00--13:00) indicate potential windows of demand flexibility, during which consumers may strategically shift electricity usage to optimise both economic outcomes (e.g., lower prices) and decarbonisation targets \citep{BOKDE2021}.

\subsection{Literature review}

In the CO$_2$ forecasting framework, a distinction can be made between short-term and long-term CO$_2$ emissions forecasting. Long-term forecasting, e.g. annual projections, enables policymakers to plan and manage future carbon footprints. By contrast, short-term CO$_2$ emissions forecasting is essential for scheduling electricity production and consumption in ways that minimise resulting emissions over a day-ahead horizon. Although long-term forecasting has received substantial attention in the literature, studies focused on short-term forecasting are limited \citep{BOKDE2021,LEERBECK2020}. \cite{FINENKO2016} analyse hourly CO$_2$ emissions from electricity generation in Singapore, highlighting hourly patterns due to both supply-side factors (e.g., renewable integration) and demand-side dynamics (e.g., peak vs. off-peak loads). \cite{MASON2018} applied evolutionary neural networks to forecast short-term CO$_2$ emissions in Ireland over a multi-hour horizon. \cite{Lowry2018} forecasted the CO$_2$ intensity of the United Kingdom's electricity grid on a day-ahead horizon using seasonal autoregressive integrated moving average (SARIMA) models and artificial neural networks. \cite{LEERBECK2020} propose a machine learning algorithm to forecast short-term average and marginal CO$_2$ emissions intensities in the Danish bidding zone DK2. The proposed algorithm combines multiple linear regression models through Softmax-weighted averaging, with residuals corrected using a SARIMA model with exogenous variables (SARIMAX). \cite{BOKDE2021} introduced two time series decomposition methods to forecast short-term CO$_2$ emissions from electricity generation over the 48-hour horizon. The proposed methods are also compared with SARIMA models, model-based on differenced pattern sequences, feed-forward neural networks, pattern sequence-based forecasting methods, and support vector machine. Short-term CO$_2$ emissions forecasts are obtained for five European countries: France, Germany, Poland, Denmark, and Norway. \cite{Maji2022b} introduced the day-ahead carbon forecasting system, which forecasts the electricity generation of each source through an artificial neural network, then integrates these forecasts with the emission rates associated with each source to calculate the overall carbon emissions and carbon intensity. \cite{Maji2022a} introduced CarbonCast, a machine learning system to forecast multi-day carbon intensities. CarbonCast has a hierarchical design that combine neural networks, weather forecasts, and historical data. Many of these studies \citep{Lowry2018,LEERBECK2020,Maji2022b,ElectricityMap2022} provide forecasts for the day ahead, while some others provide multi-day forecasts \citep{MASON2018, BOKDE2021, Maji2022a}. 

\cite{Jin2024} provided a comprehensive review of carbon emission prediction models, identifying four main categories\footnote{This classification closely aligns with that identified by \cite{Damon2002}, who grouped models into three main categories: (i) time series and regression models-including ARMA models, threshold autoregressive models, generalized autoregressive conditional heteroskedasticity (GARCH) models, non-parametric regression models, generalized additive models, and adaptive regression splines; (ii) neural networks; and (iii) classification and regression tree models.}\textsuperscript{,}\footnote{Another interesting literature review on carbon emission forecast and model classification is provided by \cite{su17041471}.}: statistical models, shallow intelligent models, neural networks, and combined models.

Statistical models are the most used because of their interpretability and reliance on historical data. Among them, the grey model is particularly popular for scenarios with limited data \citep{deng1982}. Regression models, such as linear and panel regression, have been used to analyse trade-related emissions and regional trends \citep{SINGH2015271}. Furthermore, time series models such as SARIMA and autoregressive distributed lag models \citep{lin2019assessing} have proven effective in capturing temporal dynamics, offering good forecasting performance in various applications \citep{LI2018337,Malik2020,Kaur2023}. Despite their strengths in modelling trends and seasonality, these models may struggle with abrupt fluctuations caused by external factors.

Shallow intelligent models, such as support vector machines \citep{wen2020,Wang2020} and decision trees \citep{cui2021}, improve predictive accuracy by accommodating complex data patterns \citep{AHMAD2014102}. Meanwhile, both feed-forward and recurrent neural networks have gained attention due to their ability to model nonlinearity and time-dependent structures within emission data \citep{wen20202,Ghalandari2021,daniyal2022}.

The most recent innovations involve combined prediction models, which strategically integrate multiple individual models into a more robust and unified prediction framework \citep{TASCIKARAOGLU2014,WANG2017600}. These hybrid approaches can be further categorised into combinations of statistical-statistical \citep{cui2018,MENG2014673}, statistical-intelligent \citep{wang2018,zhao2018} and intelligent-intelligent \citep{ACHEAMPONG2019,LiYan2020}. Each configuration offers distinct advantages in terms of accuracy, interpretability, and computational efficiency, making combined models a promising direction in emission forecasting.

\subsection{Study contribution}
This study aims to extend the comparison of statistical models for a one-day-ahead forecast of CO$_2$ emissions in the Italian electricity market over the period 2021-2023. The analysis is carried out at both the aggregate national level and across the seven electricity market zones: North, Centre-North, South, Centre-South, Calabria, Sicily and Sardinia.

Several model classes are explored. First, classical linear parametric models SARIMAX. Second, functional parametric models, such as the seasonal functional autoregressive (S)FAR and the functional autoregressive model with exogenous inputs (SFARX). Third, (semi) non-parametric, possibly non-linear, models, including generalised additive models (GAM) and the trigonometric seasonality, Box-Cox transformation, ARMA errors, trend, and seasonal (TBATS) modelling approach. Fourth, a "semi-functional" model is also considered as the K-nearest neighbours (KNN). 
All models are compared with a Na\"ive benchmark, in which the best prediction for a given hour is simply the value observed at the same hour the day before.

The comparison is further extended by evaluating forecast combination techniques that aggregate the predictions from the individual models described above. Specifically, combination methods include the simple average, \cite{BG1969}, and a selection-based combination strategy.

Forecast performance is evaluated using the hourly average root mean square error (RMSE), the out-of-sample $R^2$, the Diebold-Mariano (DM) test and the model confidence set (MCS).

The rest of the paper is structured as follows: CO$_2$ emissions from electricity generation are described in Section \ref{sec.data}; Section \ref{sec.methodology} introduces the single models (\ref{sec.sing.mod.setup}) and forecast combination methods (\ref{sec.forecast.comb.setup}); the forecast experiment and its results are presented in Section \ref{sec.results}; some implications of the policy on the scheduling of the electricity market are provided in Section \ref{sec.policy}; finally, Section \ref{sec.conclusions} provides concluding remarks.

\section{Data}\label{sec.data}

Hourly carbon emissions from electricity generation, denoted as $E_t$, are calculated using the following emission factor-based method \citep{BERTOLINI2025}
\begin{equation}
\begin{split}
    &E_t = \sum_{f=1}^{F} E_{t,f},\\
    &E_{t,f} = G_{t,f} \times EF_f \times O_f \times M,
\end{split}
\end{equation}
where:
\begin{itemize}
    \item $E_{t,f}$ denotes carbon emissions at time $t$ for fuel type $f$, $G_{t,f}$ represents the hourly electricity generation at time $t$ by fuel type $f$;
    \item $EF_f$ is the country-specific emission factor for fuel type $f$;
    \item $O_f$ is the oxidation rate for fuel type $f$;
    \item $M$ is the ratio of the molecular weight of CO$_2$ to the atomic weight of carbon fixed to $44/12=3.6667$.
\end{itemize}

Electricity generation data are sourced from the European Network of Transmission System Operators for Electricity \citep{entsoe}, and cover the 3-year period 1 January 2020 - 31 December 2023. Data are collected at both the national aggregate level and disaggregated across the seven electricity market zones: North, Centre-North, Centre-South, South, Calabria, Sicily and Sardinia. The data have an hourly frequency over 24 hours. The types of combusted fuels that generate electricity are: fossil brown coal, lignite, derived gas, natural gas, fossil hard coal, and fossil oil. Their country-specific emission factors and oxidation rates are sourced from the Italian institute for environmental protection and research \citep{ispra}.

\begin{figure}[H]
    \centering
    \includegraphics[width=1\textwidth]{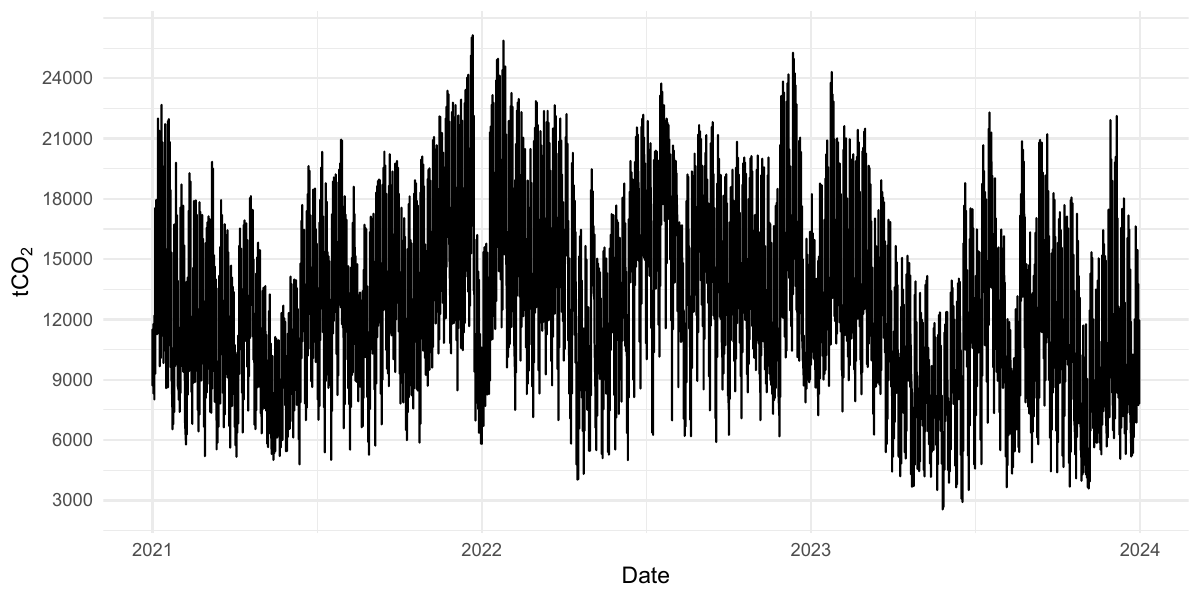}
    \caption{Carbon emissions from electricity generation in Italy (2021-2023).}
    \label{fig:carbon_emissions}
\end{figure}

Figure \ref{fig:carbon_emissions} displays the time series of CO$_2$ emissions (in tonnes of CO$_2$) from electricity generation in Italy over the period 1 January 2020 - 31 December 2023.

\begin{figure}[H]
    \centering
    \includegraphics[width=1\textwidth]{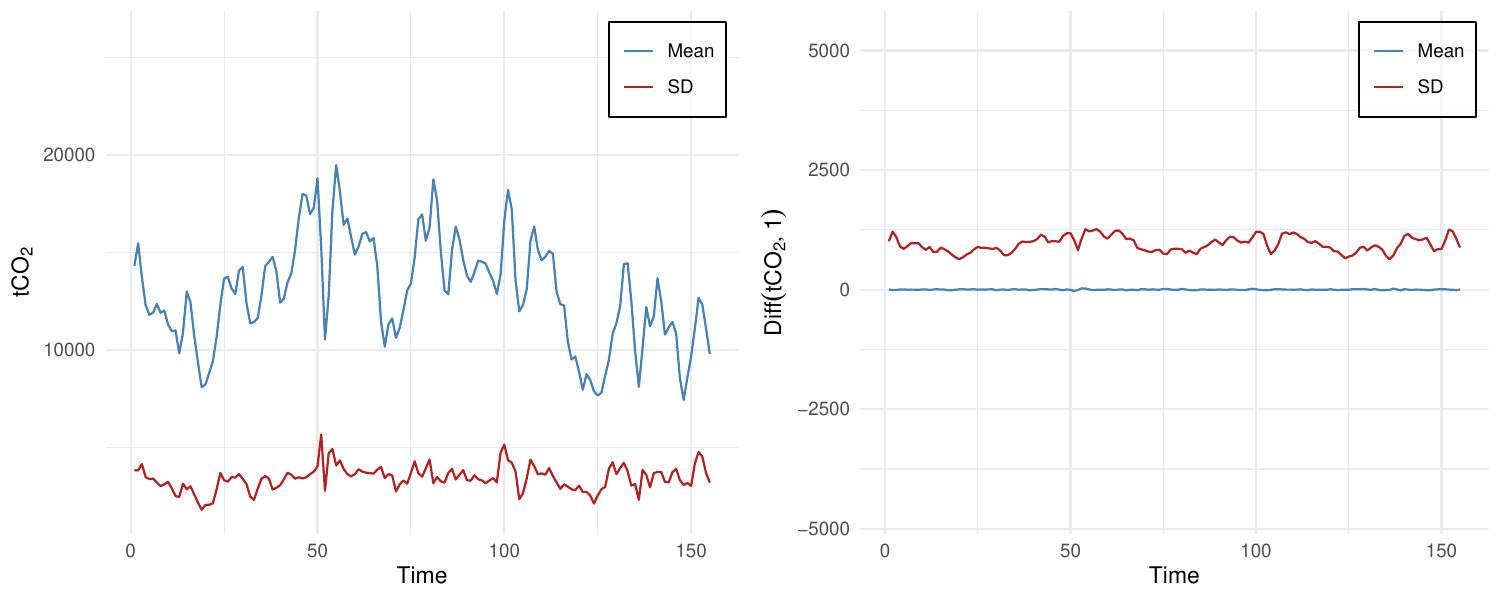}
    \caption{Two-week rolling mean and standard deviation computed with overlapping one-week windows, for the series of Italian carbon emissions (left panel) and its first differences (right panel) over the period 2021-2023.}
    \label{fig:sarimax_condition}
\end{figure}

To address potential issues regarding structural breaks and instability, Figure \ref{fig:sarimax_condition} reports the rolling mean and standard deviation. These statistics are computed using two-week windows with a one-week overlap for the series of Italian carbon emissions (left panel) and their first differences (right panel) over the period 2021-2023. The results provide no indication of structural breaks, nor of instability in the mean or variance of the differenced series. A similar pattern is also observed across the individual market zones.

\begin{figure}[H]
    \centering
    \includegraphics[width=1\textwidth]{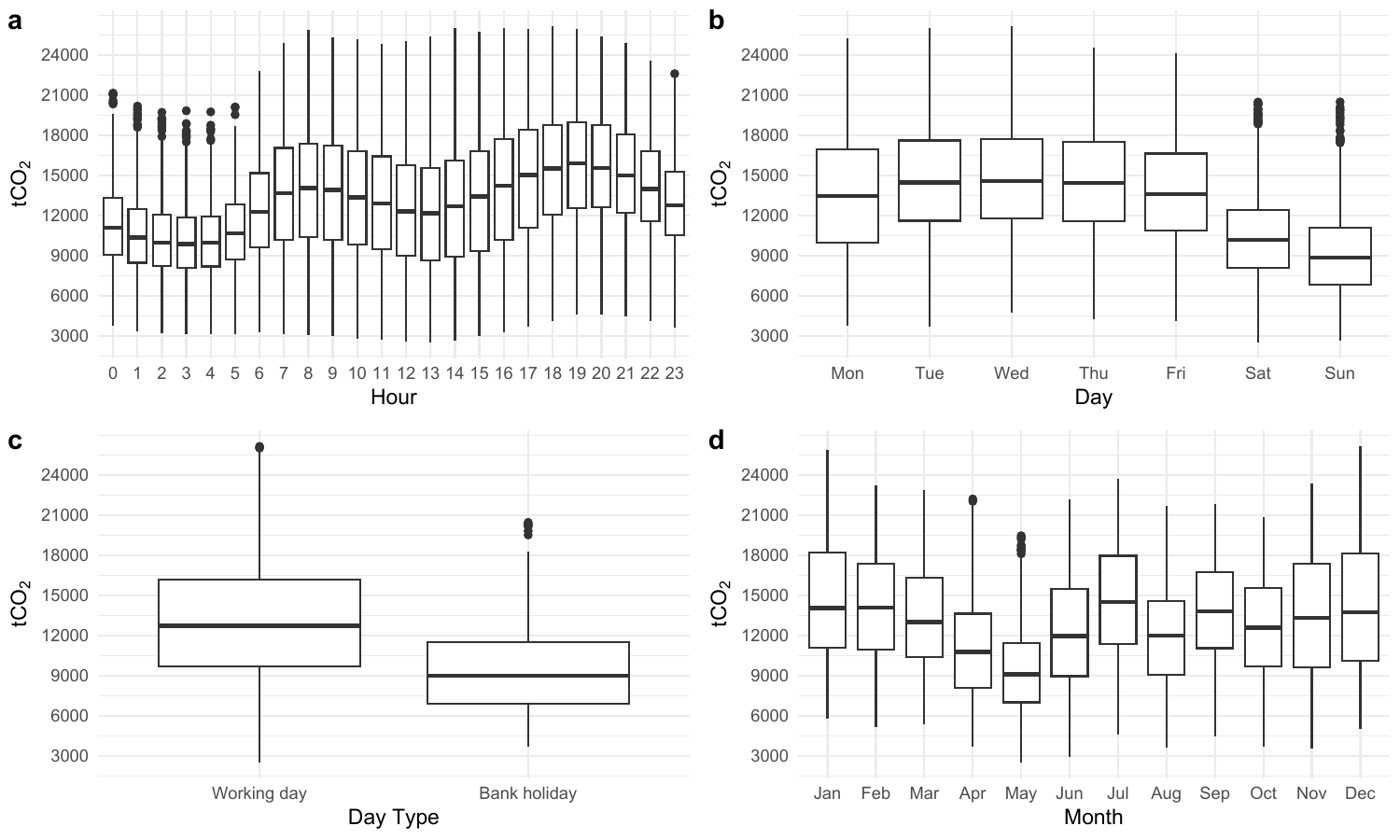}
    \caption{Boxplots of carbon emissions in Italy by seasonal components (2021-2023).}
    \label{fig:hourly_boxplot}
\end{figure}

To explore intra-period variability, Figure \ref{fig:hourly_boxplot} presents a series of box plots showing the distribution of Italian carbon emissions in different temporal dimensions: daily, weekly, bank holidays and monthly. Panel (a) highlights higher emissions during peak electricity demand hours, typically around 08:00 and 19:00, while significantly lower emissions are observed during off-peak periods, particularly between 22:00 and 05:00. Panel (b) shows higher emissions on working days (Monday to Friday), corresponding to increased industrial and commercial activity, and reduced emissions over the weekend. A similar pattern is evident in Panel (c), where emissions drop on bank holidays due to decreased demand. Panel (d) illustrates monthly fluctuations, with emissions peaking in winter due to heating needs, followed by a gradual decline in spring. A modest increase is observed in summer, primarily driven by air conditioning usage, which then decreases in August during the national holiday period, before rising again in autumn as overall energy consumption increases.

\section{Methodology}\label{sec.methodology}
This section describes the setup of single models, forecast combinations, and the overall design of the forecast experiment. In the following, the hourly observations are denoted by $E_1,\ldots, E_t, \ldots, E_n$, where $t$ indexes the hours. To indicate observation at hour $h$ of day $i$, also the notation $E_{i,h}$ is used. The relationship between $t$, $i$, and $h$ is defined by $t = 24(i - 1) + h$, with $t = 1, \ldots, 8760$, $i = 1, \ldots, 365$, and $h = 1, \ldots, 24$.

\subsection{Single model setup}\label{sec.sing.mod.setup}
The individual models used in this study are the na\"ive persistence model, the SARIMAX model, the SFARX model, the GAM, the TBATS model, and the KNN model.

To model the different seasonal components, the following calendar variables are potentially included in the various models: $H_t$ corresponds to the daily cyclical component and serves as an indicator variable for the hour at time $t$; $W_t$ denotes the weekly periodic component and is an indicator variable for the day of the week at time $t$; $B_t$ is a dummy variable for the bank holiday; $Y_t$ represents the yearly seasonal component and is an indicator variable for the specific day of the year at time $t$; $T_t$ represents the trend variable, which indicates a time index for each time instant $t$.

A na\"ive persistent model is used as a basic benchmark and is defined as follows:
\begin{equation}\label{eq.naive}
E_{t} = E_{t-24}.
\end{equation}
Here, the predicted value of carbon emissions at time $t$ corresponds to the value observed at the same hour on the previous day, that is, 24 hours earlier. The na\"ive model represents the simplest possible forecasting approach, as it does not involve parameter estimation and relies only on daily persistence.

In contrast, more complex and widely used time series models can capture the underlying temporal structures and seasonal patterns. A standard SARIMAX$(p,d,q)(P,D,Q)_S$ model is defined as follows:
\begin{equation}
\phi(B)\Phi(B^S)(1-B)^d(1-B^S)^D E_t = \mu + \theta(B)\Theta(B^S)+ \alpha_1 W_t + \alpha_2 Y_t + \alpha_3 B_t + \epsilon_t
\end{equation}
where: $B$ is the backshift operator, i.e. such that $B^k E_t = E_{t-k}$; $S$ denotes the seasonal period; $d$ and $D$ represent the orders of non-seasonal and seasonal differencing, respectively; $\phi$ is the vector of non-seasonal autoregressive coefficients; $\Phi_k$ is the vector of seasonal autoregressive coefficients; $\mu$ is a constant term; $\theta_j$ is the vector of non-seasonal moving average coefficients; and $\Theta_l$ is the vector of seasonal moving average coefficients; $\alpha_1$, $\alpha_2$ and $\alpha_3$ represent the effects of day of the week $(W_t)$, intra-annual seasonality $(Y_t)$, and bank holidays $(B_t)$, respectively; $\epsilon_t$ is a white noise error term. SARIMAX models are identified by minimising their hourly average RMSE in the calibration set. 

\begin{figure}[H]
    \centering
    \includegraphics[width=0.6\textwidth]{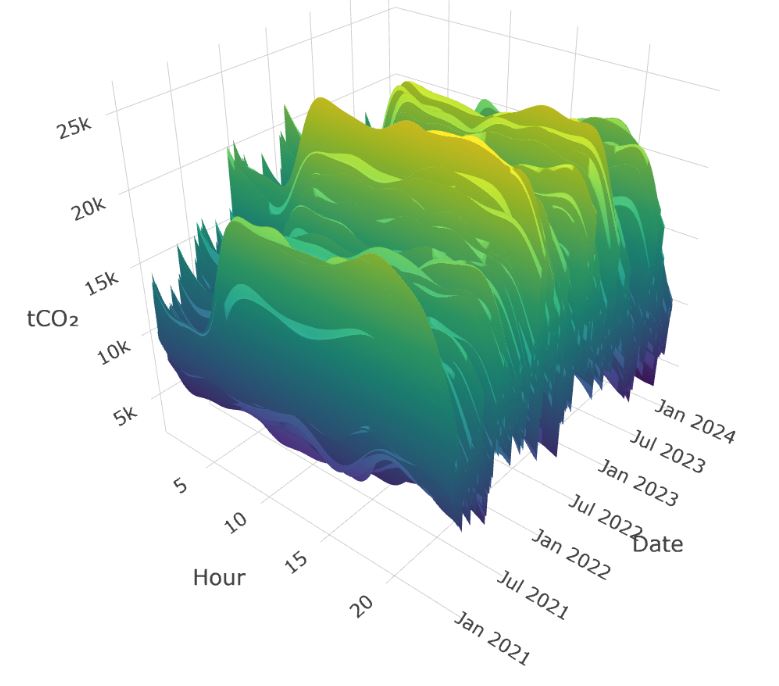}
    \caption{Daily Italian carbon emission curves (2021-2023).}
    \label{fig:fun_data}
\end{figure}

From a physical perspective, the processes involved in the production of carbon emissions from electricity generation are continuous in nature. Therefore, continuous-time stochastic processes offer a natural and potentially more accurate framework for modelling the evolution of carbon emissions. The adoption of functional models is motivated by the continuous nature of the underlying physical phenomenon \citep{ramsay2006}. The functional approach enables the recovery of information on emissions released within the intervals between discrete hourly observations, capturing intra-hour dynamics that would otherwise be overlooked. Carbon emissions from electricity generation can be naturally represented as curves, making it logical to capture the serial dependence among daily curves and their temporal dynamics \citep{Damon2002, Chen2021, Damon2024}. Figure \ref{fig:fun_data} shows the daily Italian carbon emission curves for the period 2021-2023. The raw data were transformed into functional data using the 12 Fourier basis \citep{ramsay2006}.

Within this functional approach, let $\mathcal{H}$ be a real separable Hilbert space and consider the continuous-time carbon emissions process $E_i(\tau)$ defined in $\mathcal{H}$ over a time domain $\tau\in \Tau$, i.e. $E_i(\tau)\in L^2[1,24]$ with $t \in \mathbb{N}$ indicating the $i$-th daily profile. The seasonal functional autoregressive model with exogenous input SFARX($p$) is specified as follows
\begin{equation}
E_i(\tau) =\sum_{i=1}^{p} \rho_j (E_{i-j}(\tau)) + \gamma(E_{i-7}(\tau)) + \alpha_1 (W_i(\tau)) + \alpha_2 (Y_i(\tau)) + \alpha_3 (B_i(\tau)) + \epsilon_i(\tau),
\end{equation}
where: $\rho_j(\cdot)$, $\gamma(\cdot), \alpha_1(\cdot), \alpha_2(\cdot)$ and $\alpha_3(\cdot)$ are bounded linear operators defined in $\mathcal{H}$. Specifically, the linear operator $\rho_j(\cdot)$ quantifies the influence of the lagged emission curve $E_{i-j}(\tau)$, while $\gamma(\cdot)$ quantifies the influence of the lagged emission curve of the previous week $E_{i-7}(\tau)$. The operators $\alpha_1(\cdot)$, $\alpha_2(\cdot)$, and $\alpha_3(\cdot)$ account for the effects of the weekly and yearly seasonalities and bank holidays, respectively. $W_i(\tau)$, $Y_i(\tau)$ and $B_i(\tau)$ are continuous exogenous variables defined in $\mathcal{H}$. The term $\epsilon_i(\tau)$ is a functional white noise process valued at $\mathcal{H}$. SFARX($p$) models are identified at both the aggregate level and for each zone based on their hourly average RMSE in the calibration set.

Although the SFARX model accounts for the serial correlation between the functional emission curves by modelling it as a linear operator in the functional space \citep{bosq2000linear}, non-parametric approaches such as GAM are also used to capture potential non-linearities and provide greater flexibility in modelling relationships \citep{Hastie1986,hastie2015package}. The GAM is defined by
\begin{equation}\label{eq.gam}
\begin{split}
E_t =& f_1(T_t,\lambda_1) + f_2(Y_t,\lambda_2) + f_3(H_t,\lambda_3) + f_4(W_t,\lambda_4) + f_5(E_{t-168},\lambda_5)\\ &+ f_6(E_{t-24},\lambda_6) + f_7(E_{t-48},\lambda_7) + f_8(E_{t-72},\lambda_8) + \alpha_1 B_t + \epsilon_t, 
\end{split}
\end{equation}
$f_i(\cdot,\lambda_i)$ are smoothing spline functions with $\lambda_i$ degrees of freedom. In particular, $f_1(T_t,\lambda_1=5)$ is a smoothing spline function with 5 degrees of freedom that captures the long-term trend in carbon emissions over time; $f_2(Y_t,\lambda_2=5)$ is a periodic smoothing spline function with 5 degrees of freedom that models the monthly seasonal variation; $f_3(H_t,\lambda_3)$ accounts for intra-day periodicity; $f_4(W_t,\lambda_4)$ and $f_5(E_{t-168},\lambda_5)$ capture the deterministic and stochastic components of the weekly periodicity, respectively; $f_6(E_{t-24},\lambda_6)$, $f_7(E_{t-48},\lambda_7)$, and $f_8(E_{t-72},\lambda_8)$ account for autoregressive effects at 24, 48, and 72 hours (i.e., one to three days prior); $\alpha_1 B_t$ represents the effect of bank holidays. To refer to model (\ref{eq.gam}) and its more parsimonious specifications, the (non-standard) notation GAM($p$,$P$) is used, where $p$ and $P$ represent the number of daily and weekly lags included in the model, respectively. All other variables are always present. With this notation, model (\ref{eq.gam}) is known as GAM(3,1). More parsimonious GAM specifications have been considered and selected according to the hourly average RMSE.

The TBATS model, introduced by \cite{DeLivera2011}, is another popular and flexible time series forecasting method, designed to handle multiple seasonalities, trend components, and short-term dependencies \citep{Tawiah2023,stanislaus2024}. The TBATS model \citep{Hyndman2008,Hyndman2025} can be mathematically represented as follows:
\begin{equation}\label{eq.tbats}
    E_t^{(\omega)} = l_{t-1} + \phi b_{t-1} + \sum_{i=1}^{M} S^{(i)}_{t-m_i} + d_t,
    \text{ with }E_t^{(\omega)} =
    \begin{cases} 
    \frac{E_t^\omega - 1}{\omega}, & \text{if } \omega \neq 0 \\  
    \log(E_t), & \text{if } \omega = 0  
    \end{cases}
\end{equation}
where: $E_t^{(\omega)}$ is the Box-Cox transformation of $E_t$ with parameter $\omega$; $l_t = l_{t-1} + \phi b_{t-1} + \alpha d_t$ is the local level component; $b_t = (1 - \phi) b + \phi b_{t-1} + \beta d_t$ is the short-term trend component; $\phi$ is the trend damping parameter; $S_t^{(i)}$ denotes the $i$th seasonal component at time $t$ modelled with a Fourier-based trigonometric representation; $d_t$ is the error term which follows an ARMA($p,q$) process; $m_i$ denotes the seasonal periods.

\begin{figure}[H]
    \centering
    \includegraphics[width=1\textwidth]{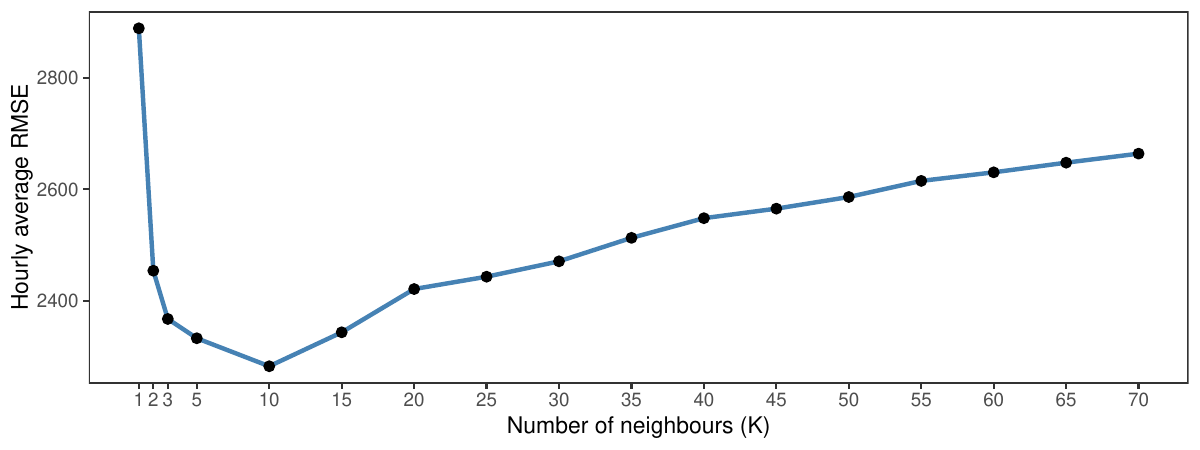}
    \caption{Identification of the optimal number of neighbours for the KNN model in the calibration set for Italy.}
    \label{fig:Elbowplot}
\end{figure}

Alongside TBATS, a "semi-functional" approach such as K-nearest neighbours (KNN) is also considered for its flexibility in leveraging similarity across daily carbon emissions profiles \citep{Lora2007,Tajmouati2024}. Given the carbon emissions recorded up to day $i-1$, the goal is to predict the daily emission profile for day $i$. Let $\mathbf{E}_i = (E_{i,1},\ldots, E_{i,h}, \ldots, E_{i,24})$ denote the vector of hourly carbon emissions for a generic day $i$, where $E_{i,h}$ is the emission recorded at hour $h$ of day $i$. Let $D(i,j)$ define the discrepancy between the daily emission profiles of days $i$ and $j$. $D(i,j)$ is evaluated using the Euclidean distance as follows:
\begin{equation}
\label{eq:Edist}
D(i, j)=\text{dist}(i, j) = \left\lVert \mathbf{E}_i - \mathbf{E}_j \right\rVert_2 = \sqrt{\sum_{h=1}^{24} \left(E_{i,h} - E_{j,h}\right)^2}.
\end{equation}
The KNN method identifies the $k$ nearest neighbours of day $i-1$, forming the neighbour set $NS = \{\mathbf{E}^{(1)},\mathbf{E}^{(2)}, \ldots, \mathbf{E}^{(k)}\}$, where $\mathbf{E}^{(1)}$ is the nearest day in the past, $\mathbf{E}^{(2)}$ the second nearest, and $\mathbf{E}^{(k)}$ the k-th nearest. The 24-hour emission profile for day $i$ is then predicted as a weighted average of the emissions on the days following those of the neighbour set:
\begin{equation}
\mathbf{E}_t = \frac{1}{\sum_{k \in NS} w_k} \sum_{k \in NS} w_k \mathbf{E}^{(k)},
\end{equation}
where $w_k \in [0,1]$ is a weight inversely proportional to the distance between $\mathbf{E}^{(k)}$ and $\mathbf{E}_{t-1}$. The optimal $k$ is chosen by minimising the hourly average RMSE \citep{HNIN2024}. An example is provided by the elbow plot in Figure \ref{fig:Elbowplot} for Italy.

\subsection{Forecast combination setup}\label{sec.forecast.comb.setup}
This section describes the forecast combination methods employed in this work.

The idea of combining forecasts dates back to \cite{BG1969}, who demonstrated that forecast combinations typically outperform individual models \citep{CLEMEN1989, surowiecki2005, TIMMERMANN2006}. Basically, different forecast combinations differ for weights given to individual models; see \cite{WANG2023} for an extensive review of the literature on forecast combinations.

Combined forecasts of carbon emissions, denoted by $\tilde{E}_{t}$, can be obtained as a linear combination of individual forecasts, formally defined as
\begin{equation}
\tilde{E}_{t} = \mathbf{w}_{t}' \mathbf{E}_{t},
\end{equation}
where $\mathbf{w}_{t} = (w_{t,1}, \dots, w_{t,N})'$ is an $N$-dimensional vector of combination weights assigned to the $N$ individual forecasts, and $\mathbf{E}_{t}$ is a $T \times N$ matrix containing the individual forecasts produced by the $N$ models. 

The simple average combination method assigns equal weights to all individual forecasts
\[
\mathbf{w}_{t}^{\text{AVE}} = \left( w_{t,1}, \dots, w_{t,N} \right)^{'}, \quad \text{where } w_{t,n} = \frac{1}{N}, \text{ for } n = 1, \dots, N.
\]
Although the equally weighted average does not consider correlations between forecast errors or the historical performance of individual forecasts, the literature suggests that it performs remarkably well compared to more complex combination techniques, becoming the most widely used combination rule \citep{BUNN1985151,GENRE2013108}.

Despite the effectiveness of the simple average, it is reasonable to assign a greater weight to the most accurate forecasting methods. \cite{BG1969} introduced a method to determine the optimal weights by minimising the variance of the combined forecast error, assigning weights inversely proportional to the variance of the forecast error, given by 
\begin{equation}
\mathbf{w}_{t}^{\text{BG}} = \frac{\boldsymbol{\Sigma}_{t}^{-1} \mathbf{1}}{\mathbf{1}' \boldsymbol{\Sigma}_{t}^{-1} \mathbf{1}},
\end{equation}  
where $\boldsymbol{\Sigma}_{T+\tau|T}$ is the covariance matrix $N \times N$ of the $t$-step forecast errors.

In addition to conventional methods, a selection-based combination approach can be employed, where the model with the lowest hourly forecast error measure \(\mathcal{E}_{h,i}\) is chosen for each forecast horizon \(t\). The best-performing model is then assigned a weight of 1, while all others receive a weight of 0. This selection-based combination method can be expressed as follows:
\[
\begin{aligned}
\mathbf{w}_{t}^{\text{SEL}} &= \mathbf{e}_{j^*}, \\
j^* &= \arg\min_{i \in \{1, \dots, N\}} \mathcal{E}_{h,i},
\end{aligned}
\]
where \(\mathbf{e}_{j^*}\) is a binary vector with a 1 at the \(j^*\) th position, indicating the selected model, and 0 elsewhere. This approach allows different models to be selected at different times of the day.

%\begin{algorithm}[H]
%\caption{Selection-Based Forecast Combination}
%\label{alg:selection}
%\begin{algorithmic}[1]
%\Require Error measures $\mathcal{E}_{h,1}, \dots, \mathcal{E}_{h,N}$
%\Ensure Weight vector $\mathbf{w}_t^{\text{SEL}}$
%\State Initialize $\mathbf{w}_t^{\text{SEL}} \gets \mathbf{0}_N$
%\State Compute the selected model index
%\State $j^* = \arg\min_{i \in \{1, \dots, N\}} \mathcal{E}_{h,i}$
%\State Assign weight 1 to the selected model
%\State $\mathbf{w}_t^{\text{SEL}}[j^*] \gets 1$
%\State \Return $\mathbf{w}_t^{\text{SEL}}$
%\end{algorithmic}
%\end{algorithm}

\section{Results}\label{sec.results}
\begin{figure}[H]
    \centering
    \includegraphics[width=0.6\textwidth]{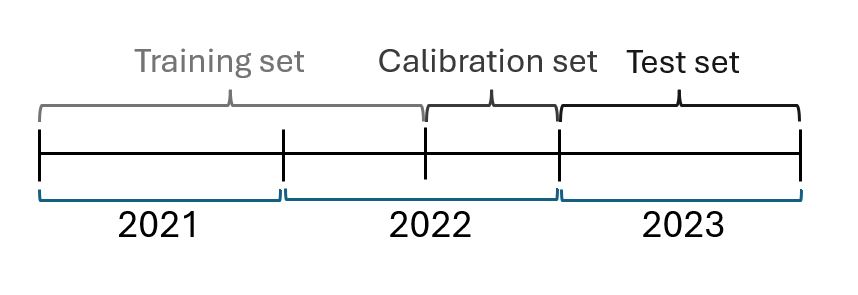}
    \caption{Forecast setup.}
    \label{fig:forecast_setup}
\end{figure}

The setup of the forecasting experiment performed in this work is shown in Figure \ref{fig:forecast_setup}. The data set is partitioned into three subsets: training, calibration, and test. The training set spans from 1 January 2021 to 30 June 2022, the calibration set from 1 July 2022 to 31 December 2022, and the test set from 1 January 2023 to 31 December 2023. 

Table \ref{tab:descriptive_stat} in the Appendix \ref{app:Tables} reports summary statistics of carbon emissions by Italian electricity market zone for training, calibration, and test sets. At the aggregate level, average daily emissions in Italy range from 12,924.16 tCO$_2$ (test set: January--December 2023) to 14,985.72 tCO$_2$ (calibration set: July--December 2022). In line with Figure \ref{fig:carbon_emissions}, a slight decrease in average emissions can be observed during 2023 compared to previous periods. At the zonal level, the North zone consistently records the highest mean emissions across all periods, ranging from 6,114.24 tCO$_2$ (test set: January--December 2023) to 6,379.07 tCO$_2$ (calibration set: July--December 2022), thus contributing substantially to national totals. Sicily and Sardinia show moderate average emission levels, characterised by relatively low variability, while the Centre-South and the South zones report higher average emissions than Sicily, but consistently lower than those of the North. In terms of variability (Stdev) there are no notable differences in the three periods.

The models are trained using data from 1 January 2021 to 30 June 2022. Following the training set, one-day-ahead forecasts are produced for the calibration period, which spans from 1 July 2022 to 31 December 2022. All models are updated daily (every 24 observations). The calibration set is used to identify the best performing model within each model class, thereby fixing the model specifications prior to evaluation on the test set. This procedure also serves to assess whether the forecasting performance observed during calibration generalises to out-of-sample data. The selection of models is based on the lowest hourly average RMSE within each class. Finally, the selected models and their forecast combinations are evaluated over the test set using forecasts with one day ahead for the period from 1 January 2023 to 31 December 2023.

Forecast performance is evaluated using the RMSE, computed separately for each hour as follows:
\begin{equation}
\text{RMSE}_h = \sqrt{\frac{1}{T_h} \sum_{i=1}^{T_h} (E_{i,h} - \hat{E}_{i,h})^2},
\end{equation}
where $E_{i,h}$ and $\hat{E}_{i,h}$ denote the observed and predicted values, respectively, for day $i$ at hour $h$, and $T_h$ is the number of observations for hour $h$ during the period out of sample. An overall measure of forecast accuracy is given by the average hourly RMSE:
\begin{equation}
\text{RMSE} = \frac{1}{24} \sum_{h=1}^{24} \text{RMSE}_h.
\end{equation}

The hourly out-of-sample $R^2$ and the hourly average $R^2$ are also used to evaluate the forecast performance. The out-of-sample $R^2$ for each hour $h$ is defined as:
\begin{equation}
R^2_h = 1 - \frac{\sum_{t=1}^{T_h} (E_{i,h} - \hat{E}_{i,h})^2}{\sum_{t=1}^{T_h} (E_{i,h} - \bar{E}_h)^2},
\end{equation}
where $\bar{E}_h$ is the mean of the observed values at hour $h$ during the out-of-sample period. The hourly average out-of-sample $R^2$ is then computed as:
\begin{equation}
R^2 = \frac{1}{24} \sum_{h=1}^{24} R^2_{\text{h}}.
\end{equation}

The DM test is used to assess the statistical significance of differences in forecast accuracy between individual models and forecast combinations \citep{Diebold1995}. This test provides a formal framework to compare the predictive performance of two competing models based on their forecast error differentials. The hypothesis system of the one-sided DM test is defined as follows:
\[
\begin{cases}
H_0: \text{Models $i$ and $j$ have the same forecast accuracy;} \\
H_1: \text{Model $j$ is more accurate than model $i$ (i.e. it has a significantly higher forecast error).}
\end{cases}
\]
The DM test statistic is calculated as
\begin{equation}\label{eq.dmtest}
DM = \frac{\bar{d}}{\sqrt{(1/T) \sum_{k=-m}^{m} \gamma_k}}
\end{equation}
where: $\bar{d}$ is the sample mean of the loss differential $d_{t,ij}$, defined as $d_{t,ij} = (e_{i,t} - e_{j,t})^2$ with $e_{i,t}$ and $e_{j,t}$ denoting the forecast errors of the models $i$ and $j$ at time $t$. The loss function $L(\cdot)$ is the squared error; $\gamma_k$ denotes the $k$-th order autocovariance of the loss differential series $d_{t,ij}$; $T$ is the number of forecast periods; $m$ is the maximum lag order used to estimate the long-run variance of $d_{t,ij}$. Under the null hypothesis $H_0$, the DM statistic asymptotically follows a standard normal distribution: $DM \sim \mathcal{N}(0,1)$.

The model confidence set procedure \citep{Hansen2011} identifies the set of models whose predictive performance cannot be statistically distinguished at a given confidence level. Starting from an initial set of models $M_0$ of size $m$, the MCS determines a smaller set $\hat{M}_{1-\alpha}^*$, called the superior set of models, in which the null hypothesis of equal predictive ability is not rejected. The procedure relies on the loss differential $d_{t,ij}$ and its sample mean $\bar{d}_{ij}$. The hypothesis system can be formulated pairwise as
\[
\begin{cases}
H_0: d_{t,ij} = 0 \quad \forall i,j \\
H_1: d_{t,ij} \neq 0 \text{ for some } i,j.
\end{cases}
\]
The test statistic is defined as in equation (\ref{eq.dmtest}) and for a current set of models $M$ can be aggregated into $DM_{M} = \max_{i,j} |DM_{ij}|$. Since the asymptotic distribution of $DM_{M}$ is nonstandard, the p-value is obtained using a bootstrap procedure \citep{Hansen2011,bernardi2014}.

\subsection{Single model results}
\begin{figure}[H]
    \centering
    \includegraphics[width=1\textwidth]{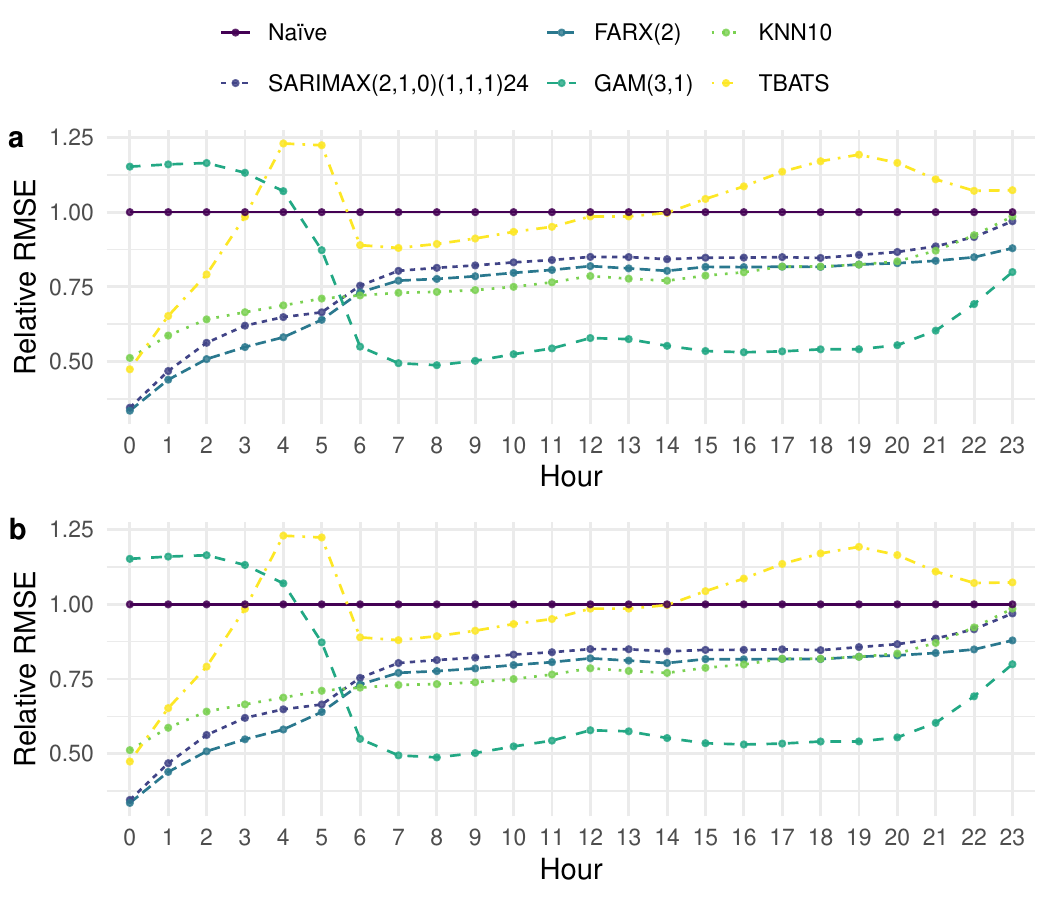}
    \caption{Relative RMSE of single models for Italy, compared to the Na\"ive model. Panel (a) calibration set and panel (b) test set.}
    \label{fig:calb_test_singm_italy}
\end{figure}

In the calibration test, the best performing models are selected by minimising the hourly average RMSE. The selected models are reported in Table \ref{tab:calb_selected_models}.\\
For Italy, the selected models are SARIMAX(2,1,0)(1,1,1)24, FARX(2), GAM(3,1), and KNN(10). Table \ref{tab:calb_singm_italy} in Appendix \ref{app:Tables} reports the hourly RMSE, hourly average RMSE and R$^2$, while the panel (a) in Figure \ref{fig:calb_test_singm_italy} displays the relative RMSE compared to the Na\"ive model. The selected GAMs for Sicily and Sardinia are GAM(2,0) and GAM(1,1), respectively. These specifications are more parsimonious compared to GAM(3,1): GAM(2,0) excludes the weekly stochastic component and incorporates autoregressive effects up to 2 days, while GAM(1,1) includes the weekly stochastic component but limits autoregressive effects to 1 day. 
\begin{table}[H]
\caption{Selected models for each model class and zone according to the hourly average RMSE and R$^2$.}
\label{tab:calb_selected_models}
\centering
\resizebox{12cm}{!}{
\begin{tabular}{lcccc}
\hline
Zone & SARIMAX & SFARX & GAM  & KNN \\
\hline
Italy &  SARIMAX(2,1,0)(1,1,1)24 & FARX(2) & GAM(3,1)  & KNN(10) \\
North & SARIMAX(3,1,0)(1,1,1)24 & FAR(2) & GAM(3,1) & KNN(10)\\
Centre North & SARIMAX(1,1,0)(1,1,0)24 & FARX(2) & GAM(3,1) & KNN(35)\\
Centre South & SARIMAX(3,1,0)(1,1,1)24 & FARX(1) & GAM(3,1) & KNN(10)\\
South & SARIMAX(1,1,0)(1,1,1)24 & FARX(3) & GAM(3,1) & KNN(10)\\
Calabria & SARIMAX(3,1,0)(1,1,1)24 & FAR(2) & GAM(3,1) & KNN(15)\\
Sicily & SARIMAX(2,1,0)(1,1,1)24 & FARX(3) & GAM(2,0) & KNN(5)\\
Sardinia & SARIMAX(1,1,0)(1,1,1)24 & FARX(1) & GAM(1,1) & KNN(10)\\
\hline
\end{tabular}
}
\end{table}

\begin{figure}[H]
    \centering
    \includegraphics[width=1\textwidth]{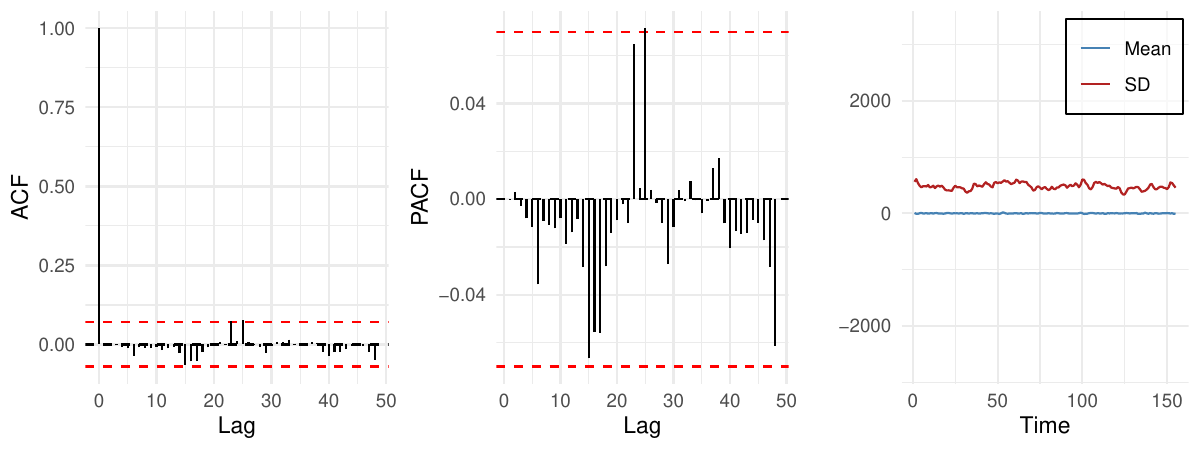}
    \caption{Residual diagnostics for the SARIMAX(2,1,0)(1,1,1)24 model estimated over the whole 2021-2023 period for Italy. The left panel displays the ACF of the residuals, the central panel reports the PACF, and the right panel shows the rolling mean and standard deviation computed with one week overlap window.}
    \label{fig:sarimax_res_check}
\end{figure}

For the test set, Table \ref{tab:test_sing_models_italy} in Appendix \ref{app:Figures} reports the forecast performance of the best individual model of each modelling approach for Italy. RMSE values are provided for each hour of the day (0 to 23), along with the hourly average RMSE and the hourly average out of sample $R^2$. The GAM(3,1) model produces the lowest RMSE for most hours, especially during the daytime hours from 6:00 to 23:00. In contrast, the FARX(2) model performs best in the early morning hours from 0:00 to 5:00. The Na\"ive and TBATS models generally exhibit higher RMSE values. SARIMAX(2,1,0)(1,1,1)24 and KNN(10) models show moderate performance and remain behind the FARX(2) and GAM(3,1) models.\\
Figure \ref{fig:sarimax_res_check} presents the residual diagnostics for the SARIMAX(2,1,0)(1,1,1)24 model estimated over the whole 2021-2023 period for Italy. The ACF and PACF (left and central panels) indicate that autocorrelations and partial autocorrelations remain within $\pm0.07$ over the first 48 lags. The panel on the right displays the rolling mean and standard deviation, computed with one week overlapping window, showing no evidence of structural breaks over the sample. Furthermore, a Lagrange Multiplier test is performed to assess the presence of residual heteroscedasticity. The test yields a statistic of 20.552 with a p-value of 0.1137. The null hypothesis of no autoregressive conditional heteroscedasticity effects cannot be rejected. Overall, the residual analysis suggests that the weak stationarity assumption is satisfied, with residuals that exhibit no significant autocorrelation and approximately constant variance. The other market zones exhibit similar diagnostic results.

The hourly average RMSE and out-of-sample $R^2$ further confirm that the GAM(3,1) model is the best, followed by the FARX(2) and KNN(10) models. Although the GAM(3,1) model provides the most accurate and reliable forecasts during daytime hours, the FARX(2) model may be preferable in the early morning. These findings are further supported by the panel (b) in Figure \ref{fig:calb_test_singm_italy}, which displays the relative RMSE compared to the Na\"ive model for Italy in the test set.

Figures \ref{fig:test_singm_marketzones}-\ref{fig:test_singm_marketzones2} in Appendix \ref{app:Figures} show the hourly performance of the best models across the seven market zones. Overall, GAM(3,1) consistently delivers the best performance in the North, Centre-North, Centre-South, South, and Calabria zones. In Sicily, FARX(3) is the best model, while in Sardinia, FARX(1) performs best, with only minor differences among the other models.

\subsection{Forecast combination results}
Based on the information obtained during the calibration process, forecast combinations can be constructed to enhance predictive performance. Although numerous alternatives were explored, only the most relevant forecast combinations are retained for analysis. These are reported in Table \ref{tab:forecast_comb_cons}. 

\begin{table}[H]
\caption{Forecast combinations.}
\label{tab:forecast_comb_cons}
\centering
\resizebox{10cm}{!}{
\begin{tabular}{llllll}
\hline
\multicolumn{2}{l}{Model} & \multicolumn{2}{l}{Method} & \multicolumn{2}{l}{Combination}\\
\hline
\multicolumn{2}{l}{\multirow{2}{*}{\begin{tabular}[c]{@{}l@{}}COMB1                                                  \\ \textit{SARIMAX, SFARX, GAM, TBATS, KNN}\end{tabular}}} & \multicolumn{2}{l}{SA}  & \multicolumn{2}{l}{COMB1.SA}  \\ 
\multicolumn{2}{l}{}                                                                                                                                                   & \multicolumn{2}{l}{BG}  & \multicolumn{2}{l}{COMB1.BG}  \\ \hline
\multicolumn{2}{l}{\multirow{3}{*}{\begin{tabular}[c]{@{}l@{}}COMB2                                                       \\ \textit{SFARX, GAM}\end{tabular}}}                 & \multicolumn{2}{l}{SA}  & \multicolumn{2}{l}{COMB2.SA}  \\
\multicolumn{2}{l}{}                                                                                                                                                   & \multicolumn{2}{l}{BG}  & \multicolumn{2}{l}{COMB2.BG}  \\
\multicolumn{2}{l}{}                                                                                                                                                   & \multicolumn{2}{l}{SEL} & \multicolumn{2}{l}{COMB2.SEL} \\ \hline
\multicolumn{2}{l}{\multirow{3}{*}{\begin{tabular}[c]{@{}l@{}}COMB3                                                  \\ \textit{SARIMAX, GAM}\end{tabular}}}                    & \multicolumn{2}{l}{SA}  & \multicolumn{2}{l}{COMB3.SA}  \\ 
\multicolumn{2}{l}{} & \multicolumn{2}{l}{BG}  & \multicolumn{2}{l}{COMB3.BG}  \\ 
\multicolumn{2}{l}{} & \multicolumn{2}{l}{SEL}  & \multicolumn{2}{l}{COMB3.SEL}\\ \hline
\multicolumn{2}{l}{\multirow{3}{*}{\begin{tabular}[c]{@{}l@{}}COMB4                                                           \\ \textit{GAM, KNN}\end{tabular}}}               & \multicolumn{2}{l}{SA}  & \multicolumn{2}{l}{COMB4.SA}  \\
\multicolumn{2}{l}{}                                                                                                                                                   & \multicolumn{2}{l}{BG}  & \multicolumn{2}{l}{COMB4.BG}  \\
\multicolumn{2}{l}{}                                                                                                                                                   & \multicolumn{2}{l}{SEL} & \multicolumn{2}{l}{COMB4.SEL}\\
\hline
\end{tabular}
}
\end{table}

Each combination combines the best performing models among those listed. BG, SA and SEL denote the combination method: BG refers to \cite{BG1969}, SA denotes the simple average, and SEL indicates the selection-based combination method. For example, COMB1.BG for Italy combines the forecasts of SARIMAX(2,1,0)(1,1,1)24, FARX(2), GAM(3,1), TBATS and KNN(10), using \cite{BG1969}. In contrast, COMB2.SEL combines the FARX(2) and GAM(3,1) forecasts using the selection-based approach.

\begin{figure}[H]
    \centering
    \includegraphics[width=1\textwidth]{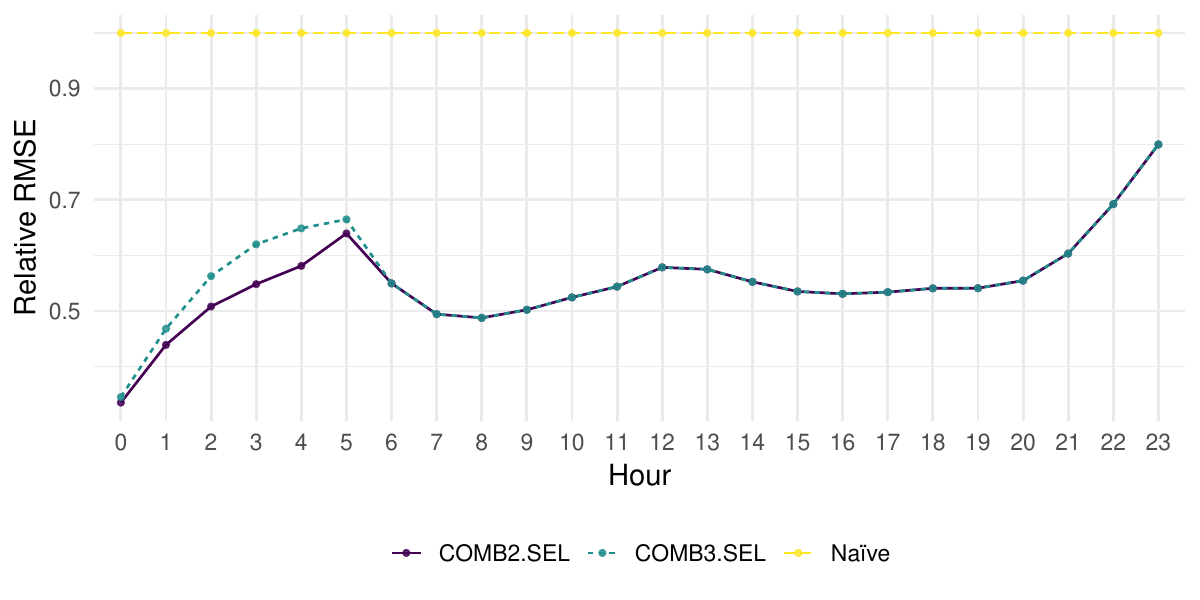}
    \caption{Italy - Relative RMSE of the two top-performing forecast combinations (based on hourly average RMSE) compared to the Na\"ive model.}
    \label{fig:test_italy_fcomb}
\end{figure}

The results for Italy are listed in Table \ref{tab:test_fcomb_italy} in the Appendix \ref{app:Tables}, which reports the hourly average RMSE of the forecast combinations in the test set. The selection-based combination method produces lower hourly average RMSE and higher out-of-sample R$^2$ compared to the simple average and \cite{BG1969}. However, simple average combination yields a lower RMSE than the \cite{BG1969} for combinations including SFARX and GAM (COMB2.SA), as well as for the combination involving all models (COMB1.SA).\\
The best-performing method is COMB2.SEL, which selects the forecasts of the FARX(2) model during the early hours of the day (00:00-05:00) and those of the GAM(3,1) model for the remaining hours (06:00-23:00). This combination reduces RMSE by 45.26\%, 15.09\%, and 19.02\% relative to the Na\"ive, GAM(3,1), and FARX(2) models, respectively.\\
The second-best combination, COMB3.SEL, achieves an hourly average RMSE of 1581.96, corresponding to a 35.21\% reduction compared to the Na\"ive benchmark. Figure \ref{fig:test_italy_fcomb} shows for Italy the relative RMSE of the two top-performing forecast combinations, compared to the Na\"ive model.\\

Table \ref{tab.DMtest} provides the DM test p-values, computed for the Italian CO$_2$ prediction, comparing the forecast accuracy of the individual models with that of the two top-performing forecast combinations based on hourly average RMSE. Each p-value compares the model in the column ($j$) with the model in the column ($i$). The null hypothesis assumes equal forecast accuracy, whereas the alternative states that the column model $j$ yields more accurate forecasts than the row model $i$. In terms of individual models, the TBATS specification is significantly less accurate than all other single models and forecast combinations. Similarly, the SARIMAX(2,1,0)(1,1,1)24 model is significantly less accurate than all competitors except the TBATS model. The FARX(2) and KNN(10) models perform better than the SARIMAX(2,1,0)(1,1,1)24 and TBATS models. By contrast, FARX(2) and KNN(10) have forecast performances that are almost statistically equivalent. The GAM(3,1) model outperforms all other individual models, although it remains less accurate than all forecast combinations. COMB2.SEL provides the most accurate forecast combination, followed by COMB4.SEL. These results are consistent with the RMSE evaluation, confirming that forecast combinations improve accuracy beyond that of the best single models.

\begin{table}[H]
    \caption{Diebold-Mariano test p-values for pairwise comparisons for Italy.}
    \label{tab.DMtest}
    \centering
    \resizebox{\textwidth}{!}{
\begin{tabular}{lccccccc}
  \hline
 & SARIMAX$^a$ & FARX(2) & GAM(3,1) & TBATS & KNN(10) & COMB2.SEL & COMB3.SEL \\ 
  \hline
SARIMAX$^a$ & - & $<0.05$ & $<0.05$ & 1.000 & $<0.05$ & $<0.05$ & $<0.05$ \\ 
  FARX(2) & 1.000 & - & $<0.05$ & 1.000 & 0.04 & $<0.05$ & $<0.05$\\ 
  GAM(3,1) & 1.000 & 1.000 & - & 1.000 & 1.000 & $<0.05$ & $<0.05$ \\ 
  TBATS & $<0.05$ & $<0.05$ & $<0.05$ & - & $<0.05$ & $<0.05$ & $<0.05$\\ 
  KNN10 & 1.000 & 0.96 & $<0.05$ & 1.000 & - & $<0.05$ & $<0.05$ \\ 
  COMB2.SEL & 1.000 & 1.000 & 1.000 & 1.000 & 1.000 & - & 1.000 \\ 
  COMB3.SEL & 1.000 & 1.000 & 1.000 & 1.000 & 1.000 & $<0.05$ & - \\ 
   \hline
\end{tabular}
    }
    \begin{tablenotes}
\item[]{\footnotesize \textit{Note}. $^a$ the model order is (2,1,0)(1,1,1)24.}
\end{tablenotes}
\end{table}

\newpage
For the North zone, findings are in line with those obtained for Italy, as the North zone accounts for the largest share of electricity generation and associated carbon emissions. The optimal combination is COMB2.SEL, with an hourly average RMSE of 930.16, which is 46.88\% lower than that of the Na\"ive model model. The second-best alternative has an RMSE that is 1.85\% higher than that of COMB2.SEL.\\
The simple average can be considered the best combination method for the other market zones producing minimal or essentially equivalent errors compared to the others combinations methods (with numerically almost identical values). For the Centre-North zone, the optimal combination merges the functional autoregressive model with GAM, yielding an RMSE 22.93\% lower than that of the Na\"ive model. By contrast, for the Centre-South, South, Calabria and Sicily, the best-performing combination pairs the SARIMAX model with GAM. These forecast combinations provide RMSE reductions of 28.54\%, 24.57\%, 25.32\% and 20.55\% relative to the Na\"ive model for the Centre-South, South, Calabria and Sicily, respectively.\\
For Sardinia, as expected, the best combination includes all models, since the individual models already perform well (Figure~\ref{fig:test_singm_marketzones2}).\\
The DM test is also performed for the market zones, comparing the individual models with the best two forecast combination. The p-values of the DM test are reported in Tables \ref{tab.DMtestnorth}--\ref{tab.DMtestsardinia} in Appendix \ref{app:Tables}. The results confirm that the North zone aligns with the overall Italian outcomes, where the selection-based combination method performs best. In the other market zones, the simple average is the most effective combination method, producing minimal or equivalent forecast performance with the others methods.

The MCS is applied to Italy and to each bidding market zone. It considers the Na\"ive model, the individual models, and the two best forecast combinations. Table \ref{tab:mcs_results} reports the superior set of models at the significance level $\alpha=0.15$, ordered according to the ranks produced by the MCS procedure.\\
For Italy and the North, the superior set includes only the forecast combination of the functional autoregressive model and the GAM. For the Centre-North, Centre-South, South, Calabria, and Sicily, the superior set includes the two best forecast combinations, which produce similar results. In Sardinia, the superior set includes the two best forecast combinations together with three individual models: FARX(1), TBATS, and SARIMAX(2,1,0)(1,1,1)24.\\
The last row of Table 4 reports the number of models excluded from the superior set for each market zone. Italy and the North exclude seven models, reflecting a more selective optimal set, while Centre-North, Centre-South, South, Calabria, and Sicily exclude six models each. In Sardinia, only three models are excluded, indicating that more models perform similarly well in this zone.\\
The MCS findings are consistent with those of the DM test. The simple average and the selection-based method perform best. Forecast combinations improve the accuracy of individual models. The most effective combinations include the functional autoregressive model and the GAM (Italy, North, Centre-North, and Calabria) or the SARIMAX model and the GAM (Centre-South, South, and Sicily). In Sardinia, where most individual models perform well, the best combination includes all models.

\begin{table}[H]
\caption{Model confidence set results. Superior set of models for $\alpha=0.15$. The MCS p-value for each model are between brackets.}
\label{tab:mcs_results}
\centering
\resizebox{\textwidth}{!}{
\begin{tabular}{cccccccccc}
\hline
Rank & Italy & North & Centre-North & Centre-South & South & Calabria & Sicily & Sardinia \\
\hline
1 & \makecell{COMB2.SEL \\ (1.00)} & \makecell{COMB2.SEL \\ (1.00)} & \makecell{COMB2.SA \\ (1.00)}  & \makecell{COMB3.SA \\ (1.00)} & \makecell{COMB3.SA \\ (1.00)} &  \makecell{COMB2.SEL \\ (1.00)} & \makecell{COMB3.SA \\ (1.00)} &  \makecell{COMB1.SA \\ (1.00)} \\
2 & - & - &\makecell{COMB2.SEL \\ (0.65)} & \makecell{COMB2.SEL \\ (0.29)} & \makecell{COMB2.SEL \\ (0.19)} & \makecell{COMB3.SA \\ (0.84)} & \makecell{COMB2.BG \\ (0.48)} & \makecell{COMB1.BG \\ (1.00)}\\
3 & -&-&-&-&-&-&-& \makecell{FARX(1) \\ (1.00)}\\
4 & -&-&-&-&-&-&-& \makecell{TBATS \\ (0.72)}\\
5 & -&-&-&-&-&-&-&  \makecell{SARIMAX$^a$ \\ (0.21)}\\
\hline
\makecell{Excl.} & 7 & 7 & 6 & 6 & 6 & 6 & 6& 3\\
\hline
\end{tabular}
}
        \begin{tablenotes}
\item[]{\footnotesize \textit{Note}. $^a$ the model order is (2,1,0)(1,1,1)24. The last row reports the number of exuded models from the superior set of models.}
\end{tablenotes}
\end{table}

\section{Policy implications on electricity market scheduling}\label{sec.policy}
The results presented in the previous sections highlight the importance of selecting an appropriate method for predicting short-term CO$_2$ emissions. Forecast performance varies across the electricity market zones, reflecting variations in local energy mixes. For instance, the outcomes for Italy and its mainland regions differ from those observed for the islands of Sicily and Sardinia. These differences are largely due to variations in energy sources and the degree of interconnection with the mainland, both of which affect electricity generation and associated CO$_2$ emissions \citep{Sapio2016,Sapio2020,Weinhold2021}. Furthermore, the use of forecast combinations, which leverage the strengths of multiple individual models, has been shown to enhance predictive accuracy in most cases, particularly when the models capture different structural characteristics of the data \citep{WANG2023}

\begin{figure}[H]
    \centering
    \includegraphics[width=1\textwidth]{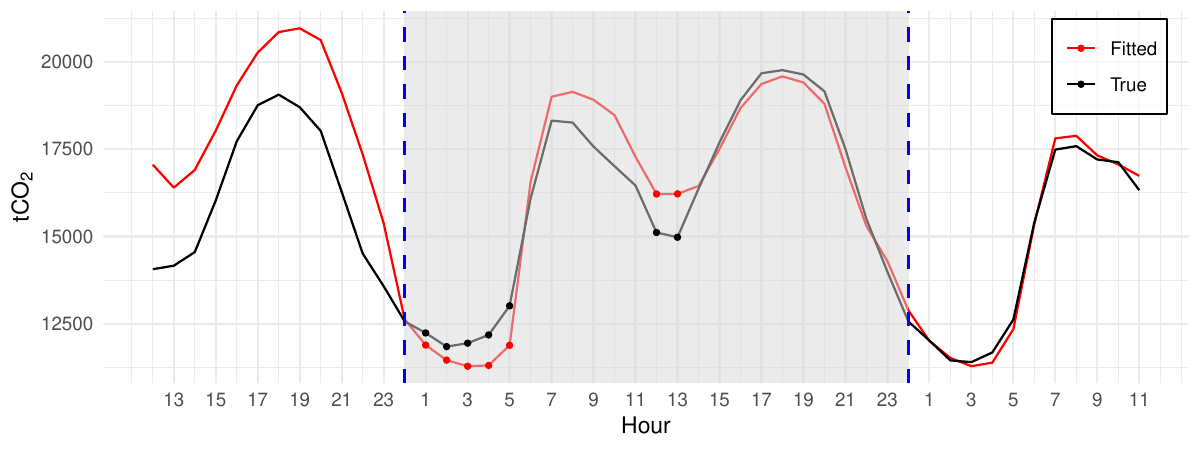}
    \caption{Example of scheduling 7 hours of flexible electricity consumption one day in advance in Italy. The red and black curves represent the fitted and true ex-post carbon emissions, respectively. The grey area indicates the day-ahead period. The minimum hours of the forecasted and true values are marked with red and black dots.}
    \label{fig:example_prediction}
\end{figure}

\begin{figure}[H]
    \centering
    \includegraphics[width=1\textwidth]{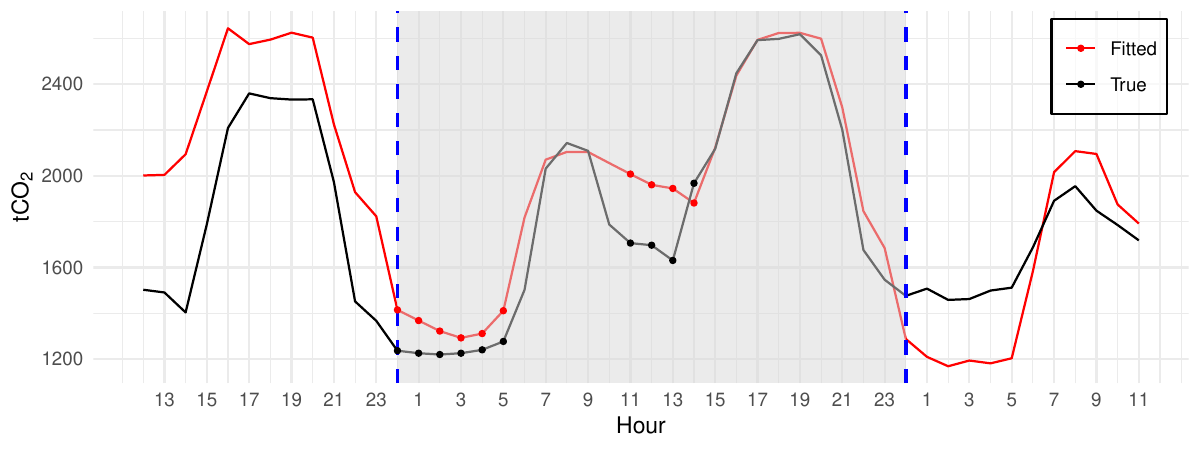}
    \caption{Example of scheduling 10 hours of flexible electricity consumption one day in advance in the Centre-South zone. The red and black curves represent the fitted and true ex-post carbon emissions, respectively. The grey area indicates the day-ahead period. The minimum hours of the forecasted and true values are marked with red and black dots.}
    \label{fig:example_prediction2}
\end{figure}

These findings highlight the importance of considering zonal specificities when implementing flexible energy demand strategies \citep{Abolghasemi2025}. Short-term CO$_2$ emissions forecasting enables flexible electricity demand scheduling to minimise emissions \citep{BOKDE2021}. For example, Figure \ref{fig:example_prediction} shows true ex-post (black) and predicted (red) carbon emissions using COMB2.SEL for Italy over 48 consecutive hours (1-3 February 2023). Mean emissions during this period are 16,345.42 tCO$_2$ and 15,712.65 tCO$_2$ for predicted and true values, respectively. The series starts at 12:00, when the day-ahead market is cleared; bids submitted before this time cover consumption for the following day (grey area between dashed blue lines). Flexible consumption is scheduled during hours with the lowest predicted emissions: 01:00-05:00 and 12:00-13:00. The red and black dots indicate the predicted and true minimum values, respectively.\\ 
Similar patterns have been reported for France, Germany, Norway, Denmark, and Poland \citep{BOKDE2021}, although the timing of flexible demand varies across countries. In Germany and France, demand is usually scheduled between 09:00 and 13:00, while in Denmark it occurs between 16:00 and 21:00. These differences underline the need to plan flexible demand according to each country and market area. Figure \ref{fig:example_prediction2} shows the scheduling of 10 hours of flexible electricity demand in the Centre-South, illustrating that the timing of flexible energy demand windows can differ from those observed for Italy in Figure \ref{fig:example_prediction}.

\section{Conclusions}\label{sec.conclusions}
This study investigates the short-term forecasting of carbon emissions from electricity generation in the Italian power market. Using hourly data from 2021 to 2023, several statistical models and forecast combination methods are evaluated and compared at the national and zonal levels. Four main model classes are considered: linear parametric models, functional parametric models, non-parametric and possibly non-linear models, and a semi-functional approach. Forecast combinations include the simple averaging, \cite{BG1969}, and the selection-based approach.

Forecast performance varies slightly between the different electricity market zones, with particularly distinct patterns observed in the Sicilian and Sardinian markets.

For mainland Italy, including the national level, North, Centre-North, Centre-South, South, and Calabria, GAM provides the most accurate forecasts during daytime hours, whereas functional parametric models perform best during the early morning period. In general, GAM emerges as the most effective individual model based on the hourly average RMSE and DM test results. Among the forecast combination approaches, the selection-based method consistently performs best in Italy and the North, while the simple average is the most effective in the remaining market zones.\\
In Italy, the North, and the Centre-North, the most effective forecast combinations are those that employ the functional autoregressive model together with GAM. In contrast, in the Centre-South, South, Calabria, and Sicily, the best-performing combinations pair the SARIMAX model with GAM.\\
In Sardinia, all individual models deliver reasonably strong forecasting performance (Figure \ref{fig:test_singm_marketzones2}). As expected, the best-performing combination is the one that incorporates all models.

This work can be expanded in several ways. First, future work can integrate other approaches and models, such as neural networks and deep learning methods \citep{Lowry2018,BOKDE2021,Maji2022b}. Second, multi-day forecasts can also be explored, which can potentially allow for more advanced demand-side management strategies \citep{Maji2022a,MASON2018,GIROLIMETTO2024a}. Third, it is worth investigating the forecast reconciliation methods \citep{Wickramasuriya2019,DIFONZO2024,girolimetto2024b}, as they provide a coherent framework to ensure consistency between forecasts at different levels of aggregation. This is particularly important in the context of electricity-related CO$_2$ emissions, where forecasts may be required at the zonal and national levels.

\bibliographystyle{apalike}
\bibliography{References.bib}

@misc{edgar2024,
  author       = {{European Commission}},
  title        = {{EDGAR Community GHG Database: Total CO$_2$ emissions by country and sector}},
  year         = {2024},
  publisher    = {Publications Office of the European Union},
  note         = {Used version: EDGAR\_2024\_GHG. Website: \url{https://edgar.jrc.ec.europa.eu/dataset_ghg60}. Acessed: May 08, 2025}
}

@article{KOCAK2023,
title = {The impact of electricity from renewable and non-renewable sources on energy poverty and greenhouse gas emissions (GHGs): Empirical evidence and policy implications},
journal = {Energy},
volume = {272},
pages = {127125},
year = {2023},
issn = {0360-5442},
doi = {https://doi.org/10.1016/j.energy.2023.127125},
author = {Emrah Kocak and Eyup Emre Ulug and Burcu Oralhan}
}

@misc{carbonbrief2025,
  author       = {{Carbon Brief}},
  title        = {Power-sector {CO}$_2$ hits all-time high in 2024 - despite record growth for clean energy},
  year         = {2025},
  url          = {https://www.carbonbrief.org/power-sector-co2-hits-all-time-high-in-2024-despite-record-growth-for-clean-energy/},
  note         = {Renewables, published: 2025-04-08}
}

@misc{UNFCCC2015,
  title        = {{Paris Agreement}},
  author       = {{United Nations}},
  year         = {2015},
  howpublished = {\url{https://unfccc.int/process-and-meetings/the-paris-agreement}}
}

@misc{SDG2015,
  title        = {{The 17 Sustainable Development Goals (SDGs)}},
  author       = {{United Nations}},
  year         = {2015},
  howpublished = {\url{https://sdgs.un.org/goals}}
}

@misc{TARGET2050,
  title        = {{Carbon neutrality by 2050: the world's most urgent mission}},
  author       = {{United Nations}},
  year         = {2020},
  howpublished = {\url{https://www.un.org/sg/en/content/sg/articles/2020-12-11/carbon-neutrality-2050-the-world%E2%80%99s-most-urgent-mission}}
}

@article{Abolghasemi2025,
title = {Improving Cross-temporal forecasts reconciliation accuracy and utility in energy market},
journal = {Applied Energy},
year = {2025},
doi = {DOI:10.48550/arXiv.2412.11153},
author = {Mahdi Abolghasemi and Daniele Girolimetto and Tommaso {Di Fonzo}}
}

@misc{IEA2019,
  author       = {{International Energy Agency}},
  title        = {Status of Power System Transformation 2019},
  year         = {2019},
  publisher    = {International Energy Agency},
  address      = {Paris},
  url          = {https://www.iea.org/reports/status-of-power-system-transformation-2019},
  note         = {Licence: CC BY 4.0}
}

@article{HUBER2021,
title = {Carbon efficient smart charging using forecasts of marginal emission factors},
journal = {Journal of Cleaner Production},
volume = {284},
pages = {124766},
year = {2021},
issn = {0959-6526},
doi = {https://doi.org/10.1016/j.jclepro.2020.124766},
author = {Julian Huber and Kai Lohmann and Marc Schmidt and Christof Weinhardt}
}

@article{JOCHEM2015,
title = {Assessing CO2 emissions of electric vehicles in Germany in 2030},
journal = {Transportation Research Part A: Policy and Practice},
volume = {78},
pages = {68-83},
year = {2015},
issn = {0965-8564},
doi = {https://doi.org/10.1016/j.tra.2015.05.007},
author = {Patrick Jochem and Sonja Babrowski and Wolf Fichtner}
}

@article{BIRESSELIOGLU2018,
title = {Electric mobility in Europe: A comprehensive review of motivators and barriers in decision making processes},
journal = {Transportation Research Part A: Policy and Practice},
volume = {109},
pages = {1-13},
year = {2018},
issn = {0965-8564},
doi = {https://doi.org/10.1016/j.tra.2018.01.017},
author = {Mehmet Efe Biresselioglu and Melike {Demirbag Kaplan} and Barbara Katharina Yilmaz}
}

@inproceedings{jung2019,
  title={Goal framing in smart charging-increasing bev users' charging flexibility with digital nudges},
  author={Jung, Dominik and Schaule, Elisabeth and Weinhardt, Christof},
  booktitle={Proceedings of the 27th European Conference on Information Systems: Information Systems for a Sharing Society, Stockholm, Sweden},
  pages={8--14},
  year={2019}
}

@article{huber2018,
  title={Waiting for the sun-can temporal flexibility in BEV charging avoid carbon emissions?},
  author={Huber, Julian and Weinhardt, Christof},
  journal={Energy Informatics},
  volume={1},
  pages={115--126},
  year={2018},
  publisher={Springer}
}

@article{WILL2016,
title = {Understanding user acceptance factors of electric vehicle smart charging},
journal = {Transportation Research Part C: Emerging Technologies},
volume = {71},
pages = {198-214},
year = {2016},
issn = {0968-090X},
doi = {https://doi.org/10.1016/j.trc.2016.07.006},
author = {Christian Will and Alexander Schuller}
}

@article{He2024,
    author = {He, Yanfeng and Guo, Shenglian and Zhou, Yanlai and Zhu, Di and Chen, Hua and Xiong, Lihua and Liu, Jie and Xu, Chong-Yu},
    title = {Boosting hydropower generation of mixed reservoirs for reducing carbon emissions by using a simulation-optimization framework},
    journal = {Hydrology Research},
    volume = {55},
    number = {2},
    pages = {144-160},
    year = {2024},
    month = {01},
    issn = {0029-1277},
    doi = {10.2166/nh.2023.181}
}

@article{HADDADIAN2015,
title = {Optimal scheduling of distributed battery storage for enhancing the security and the economics of electric power systems with emission constraints},
journal = {Electric Power Systems Research},
volume = {124},
pages = {152-159},
year = {2015},
issn = {0378-7796},
doi = {https://doi.org/10.1016/j.epsr.2015.03.002},
author = {G. Haddadian and N. Khalili and M. Khodayar and M. Shahidehpour}
}

@article{LEERBECK2020,
title = {Short-term forecasting of CO2 emission intensity in power grids by machine learning},
journal = {Applied Energy},
volume = {277},
pages = {115527},
year = {2020},
issn = {0306-2619},
doi = {https://doi.org/10.1016/j.apenergy.2020.115527},
author = {Kenneth Leerbeck and Peder Bacher and Rune Grønborg Junker and Goran Goranovi\'c and Olivier Corradi and Razgar Ebrahimy and Anna Tveit and Henrik Madsen}
}

@article{BOKDE2021,
title = {Short-term CO2 emissions forecasting based on decomposition approaches and its impact on electricity market scheduling},
journal = {Applied Energy},
volume = {281},
pages = {116061},
year = {2021},
issn = {0306-2619},
doi = {https://doi.org/10.1016/j.apenergy.2020.116061},
author = {Neeraj Dhanraj Bokde and Bo Tranberg and Gorm Bruun Andresen}
}

@techreport{wagner2002merit,
  author       = {Wagner, U. and Mauch, W. and von Roon, S.},
  title        = {Das Merit-Order-Dilemma der Emissionen},
  institution  = {Forschungsstelle f\"ur Energiewirtschaft e.V.},
  year         = {2002},
  type         = {Technical Report}
}

@article{voronin2013,
  author    = {Voronin, Sergey and Partanen, Jarmo},
  title     = {Price Forecasting in the Day-Ahead Energy Market by an Iterative Method with Separate Normal Price and Price Spike Frameworks},
  journal   = {Energies},
  year      = {2013},
  volume    = {6},
  number    = {11},
  pages     = {5897--5920},
  doi       = {10.3390/en6115897}
}

@article{FLESCHUTZ2021,
title = {The effect of price-based demand response on carbon emissions in European electricity markets: The importance of adequate carbon prices},
journal = {Applied Energy},
volume = {295},
pages = {117040},
year = {2021},
issn = {0306-2619},
doi = {https://doi.org/10.1016/j.apenergy.2021.117040},
author = {Markus Fleschutz and Markus Bohlayer and Marco Braun and Gregor Henze and Michael D. Murphy}
}

@article{FINENKO2016,
title = {Temporal CO2 emissions associated with electricity generation: Case study of Singapore},
journal = {Energy Policy},
volume = {93},
pages = {70-79},
year = {2016},
issn = {0301-4215},
doi = {https://doi.org/10.1016/j.enpol.2016.02.039},
author = {Anton Finenko and Lynette Cheah}
}

@article{MASON2018,
title = {Forecasting energy demand, wind generation and carbon dioxide emissions in Ireland using evolutionary neural networks},
journal = {Energy},
volume = {155},
pages = {705-720},
year = {2018},
issn = {0360-5442},
doi = {https://doi.org/10.1016/j.energy.2018.04.192},
author = {Karl Mason and Jim Duggan and Enda Howley}
}

@article{Lowry2018,
author = {Gordon Lowry},
title ={Day-ahead forecasting of grid carbon intensity in support of heating, ventilation and air-conditioning plant demand response decision-making to reduce carbon emissions},
journal = {Building Services Engineering Research \& Technology},
volume = {39},
number = {6},
pages = {749-760},
year = {2018},
doi = {10.1177/0143624418774738}
}

@inproceedings{Maji2022b,
author = {Maji, Diptyaroop and Sitaraman, Ramesh K. and Shenoy, Prashant},
title = {DACF: day-ahead carbon intensity forecasting of power grids using machine learning},
year = {2022},
isbn = {9781450393973},
publisher = {Association for Computing Machinery},
address = {New York, NY, USA},
url = {https://doi.org/10.1145/3538637.3538849},
doi = {10.1145/3538637.3538849},
booktitle = {Proceedings of the Thirteenth ACM International Conference on Future Energy Systems},
pages = {188-192},
numpages = {5},
location = {Virtual Event},
series = {e-Energy '22}
}

@inproceedings{Maji2022a,
author = {Maji, Diptyaroop and Shenoy, Prashant and Sitaraman, Ramesh K.},
title = {CarbonCast: multi-day forecasting of grid carbon intensity},
year = {2022},
isbn = {9781450398909},
publisher = {Association for Computing Machinery},
address = {New York, NY, USA},
doi = {10.1145/3563357.3564079},
booktitle = {Proceedings of the 9th ACM International Conference on Systems for Energy-Efficient Buildings, Cities, and Transportation},
pages = {198-207},
numpages = {10},
location = {Boston, Massachusetts},
series = {BuildSys '22}
}

@misc{ElectricityMap2022,
  author       = {{ElectricityMap}},
  title        = {ElectricityMap},
  year         = {2022},
  note         = {Retrieved July 28, 2022 from \url{https://electricitymap.org/}}
}

@article{Jin2024,
  author    = {Jin, Y. and Sharifi, A. and Li, Z. and Chen, S. and Zeng, S. and Zhao, S.},
  title     = {Carbon emission prediction models: A review},
  journal   = {Science of the Total Environment},
  volume    = {927},
  pages     = {1--20},
  year      = {2024}
}

@article{Kaur2023,
  author    = {Kaur, Jatinder and Parmar, Kulwinder Singh and Singh, Sarbjit},
  title     = {Autoregressive models in environmental forecasting time series: a theoretical and application review},
  journal   = {Environmental Science and Pollution Research},
  volume    = {30},
  number    = {8},
  pages     = {19617--19641},
  year      = {2023},
  doi       = {10.1007/s11356-023-25148-9},
  url       = {https://doi.org/10.1007/s11356-023-25148-9}
}

@article{LI2018337,
title = {Can China achieve its CO2 emissions peak by 2030?},
journal = {Ecological Indicators},
volume = {84},
pages = {337-344},
year = {2018},
issn = {1470-160X},
doi = {https://doi.org/10.1016/j.ecolind.2017.08.048},
url = {https://www.sciencedirect.com/science/article/pii/S1470160X17305381},
author = {Feifei Li and Zhe Xu and Hui Ma}
}

@article{Malik2020,
author = {Malik, Aysha and Hussain, Ejaz and Baig, Sofia and Khokhar, Muhammad Fahim},
title = {Forecasting CO2 emissions from energy consumption in Pakistan under different scenarios: The China-Pakistan Economic Corridor},
journal = {Greenhouse Gases: Science and Technology},
volume = {10},
number = {2},
pages = {380-389},
doi = {https://doi.org/10.1002/ghg.1968},
url = {https://scijournals.onlinelibrary.wiley.com/doi/abs/10.1002/ghg.1968},
eprint = {https://scijournals.onlinelibrary.wiley.com/doi/pdf/10.1002/ghg.1968},
year = {2020}
}

@article{lin2019assessing,
  title={Assessing Ghana's carbon dioxide emissions through energy consumption structure towards a sustainable development path},
  author={Lin, Boqiang and Agyeman, Stephen Duah},
  journal={Journal of Cleaner Production},
  volume={238},
  pages={117941},
  year={2019},
  publisher={Elsevier}
}

@article{deng1982,
  author  = {Deng, J. L.},
  title   = {Control problems of grey systems},
  journal = {Systems \& Control Letters},
  volume  = {1},
  number  = {5},
  pages   = {288--294},
  year    = {1982}
}

@article{SINGH2015271,
title = {Estimating future energy use and CO2 emissions of the world's cities},
journal = {Environmental Pollution},
volume = {203},
pages = {271-278},
year = {2015},
issn = {0269-7491},
doi = {https://doi.org/10.1016/j.envpol.2015.03.039},
url = {https://www.sciencedirect.com/science/article/pii/S0269749115001694},
author = {Shweta Singh and Chris Kennedy}
}

@article{wen2020,
  title={Influencing factors analysis and forecasting of residential energy-related CO2 emissions utilizing optimized support vector machine},
  author={Wen, Lei and Cao, Yang},
  journal={Journal of Cleaner Production},
  volume={250},
  pages={119492},
  year={2020},
  publisher={Elsevier}
}

@article{AHMAD2014102,
title = {A review on applications of ANN and SVM for building electrical energy consumption forecasting},
journal = {Renewable and Sustainable Energy Reviews},
volume = {33},
pages = {102-109},
year = {2014},
issn = {1364-0321},
doi = {https://doi.org/10.1016/j.rser.2014.01.069},
url = {https://www.sciencedirect.com/science/article/pii/S1364032114000914},
author = {A.S. Ahmad and M.Y. Hassan and M.P. Abdullah and H.A. Rahman and F. Hussin and H. Abdullah and R. Saidur}
}

@article{Wang2020,
  author  = {Wang, Luqi and Xue, Xiaolong and Zhao, Zebin and Wang, Yinhai and Zeng, Ziqiang},
  title   = {Finding the de-carbonization potentials in the transport sector: Application of scenario analysis with a hybrid prediction model},
  journal = {Environmental Science and Pollution Research},
  volume  = {27},
  number  = {17},
  pages   = {21762--21776},
  year    = {2020},
  doi     = {10.1007/s11356-020-08627-1},
  url     = {https://doi.org/10.1007/s11356-020-08627-1}
}

@article{cui2021,
  title={Forecasting of carbon emission in China based on gradient boosting decision tree optimized by modified whale optimization algorithm},
  author={Cui, Xiwen and E, Shaojun and Niu, Dongxiao and Chen, Bosong and Feng, Jiaqi},
  journal={Sustainability},
  volume={13},
  number={21},
  pages={12302},
  year={2021},
  publisher={MDPI}
}

@article{Ghalandari2021,
  author  = {Ghalandari, Mohammad and Forootan Fard, Habib and Komeili Birjandi, Ali and Mahariq, Ibrahim},
  title   = {Energy-related carbon dioxide emission forecasting of four European countries by employing data-driven methods},
  journal = {Journal of Thermal Analysis and Calorimetry},
  volume  = {144},
  number  = {5},
  pages   = {1999--2008},
  year    = {2021},
  doi     = {10.1007/s10973-020-10400-y},
  url     = {https://doi.org/10.1007/s10973-020-10400-y}
}

@article{daniyal2022,
  title={Comparison of Conventional Modeling Techniques with the Neural Network Autoregressive Model (NNAR): Application to COVID-19 Data},
  author={Daniyal, Muhammad and Tawiah, Kassim and Muhammadullah, Sara and Opoku-Ameyaw, Kwaku},
  journal={Journal of healthcare engineering},
  volume={2022},
  number={1},
  pages={4802743},
  year={2022},
  publisher={Wiley Online Library}
}

@article{wen20202,
  title={Forecasting CO2 emissions in Chinas commercial department, through BP neural network based on random forest and PSO},
  author={Wen, Lei and Yuan, Xiaoyu},
  journal={Science of The Total Environment},
  volume={718},
  pages={137194},
  year={2020},
  publisher={Elsevier}
}

@article{TASCIKARAOGLU2014,
title = {A review of combined approaches for prediction of short-term wind speed and power},
journal = {Renewable and Sustainable Energy Reviews},
volume = {34},
pages = {243-254},
year = {2014},
issn = {1364-0321},
doi = {https://doi.org/10.1016/j.rser.2014.03.033},
url = {https://www.sciencedirect.com/science/article/pii/S1364032114001944},
author = {A. Tascikaraoglu and M. Uzunoglu}
}

@article{WANG2017600,
title = {Forecasting Chinese carbon emissions from fossil energy consumption using non-linear grey multivariable models},
journal = {Journal of Cleaner Production},
volume = {142},
pages = {600-612},
year = {2017},
note = {Special Volume on Improving natural resource management and human health to ensure sustainable societal development based upon insights gained from working within Big Data Environments},
issn = {0959-6526},
doi = {https://doi.org/10.1016/j.jclepro.2016.08.067},
url = {https://www.sciencedirect.com/science/article/pii/S0959652616312161},
author = {Zheng-Xin Wang and De-Jun Ye}
}

@article{cui2018,
  title={Decomposition and forecasting of CO2 emissions in China's power sector based on STIRPAT model with selected PLS model and a novel hybrid PLS-Grey-Markov model},
  author={Cui, Herui and Wu, Ruirui and Zhao, Tian},
  journal={Energies},
  volume={11},
  number={11},
  pages={2985},
  year={2018},
  publisher={MDPI}
}

@article{MENG2014673,
title = {A small-sample hybrid model for forecasting energy-related CO2 emissions},
journal = {Energy},
volume = {64},
pages = {673-677},
year = {2014},
issn = {0360-5442},
doi = {https://doi.org/10.1016/j.energy.2013.10.017},
url = {https://www.sciencedirect.com/science/article/pii/S0360544213008669},
author = {Ming Meng and Dongxiao Niu and Wei Shang}
}

@article{wang2018,
  title={Can China achieve the 2020 and 2030 carbon intensity targets through energy structure adjustment?},
  author={Wang, Ying and Shang, Peipei and He, Lichun and Zhang, Yingchun and Liu, Dandan},
  journal={Energies},
  volume={11},
  number={10},
  pages={2721},
  year={2018},
  publisher={MDPI}
}

@article{zhao2018,
  title={Forecasting carbon dioxide emissions based on a hybrid of mixed data sampling regression model and back propagation neural network in the USA},
  author={Zhao, Xin and Han, Meng and Ding, Lili and Calin, Adrian Cantemir},
  journal={Environmental Science and Pollution Research},
  volume={25},
  number={3},
  pages={2899--2910},
  year={2018},
  doi={10.1007/s11356-017-0642-6},
  url={https://doi.org/10.1007/s11356-017-0642-6},
  publisher={Springer},
  issn={1614-7499}
}

@article{LiYan2020,
author = {Li, Yan},
title = {Forecasting Chinese carbon emissions based on a novel time series prediction method},
journal = {Energy Science \& Engineering},
volume = {8},
number = {7},
pages = {2274-2285},
doi = {https://doi.org/10.1002/ese3.662},
year = {2020}
}

@article{ACHEAMPONG2019,
title = {Modelling carbon emission intensity: Application of artificial neural network},
journal = {Journal of Cleaner Production},
volume = {225},
pages = {833-856},
year = {2019},
issn = {0959-6526},
doi = {https://doi.org/10.1016/j.jclepro.2019.03.352},
url = {https://www.sciencedirect.com/science/article/pii/S0959652619310686},
author = {Alex O. Acheampong and Emmanuel B. Boateng}
}

@article{BG1969,
  author  = {Bates, J. M. and Granger, C. W. J.},
  title   = {The Combination of Forecasts},
  journal = {Operations Research},
  year    = {1969},
  volume  = {20},
  number  = {4},
  pages   = {451--468}
}

@article{BERTOLINI2025,
title = {Accounting carbon emissions from electricity generation: A review and comparison of emission factor-based methods},
journal = {Applied Energy},
volume = {392},
pages = {125992},
year = {2025},
issn = {0306-2619},
doi = {https://doi.org/10.1016/j.apenergy.2025.125992},
author = {Marina Bertolini and Pierdomenico Duttilo and Francesco Lisi}
}

@misc{entsoe,
  title        = {Actual Generation per Production Type},
  author       = {{ENTSO-E}},
  year         = 2025,
  note         = {\url{https://transparency.entsoe.eu/content/static_content/Static%20content/knowledge%20base/data-views/generation/Data-view%20Actual%20Generation%20per%20Production%20Unit.html} [Accessed: (January 03, 2025)]}
}

@misc{ispra,
  title        = {Emission factors for the production and consumption of electricity in Italy},
  author       = {ISPRA},
  year         = 2025,
  note         = {\url{https://emissioni.sina.isprambiente.it} [Accessed: (January 03, 2025)]}
}

@article{Damon2002,
author = {Damon, Julien and Guillas, Serge},
title = {The inclusion of exogenous variables in functional autoregressive ozone forecasting},
journal = {Environmetrics},
volume = {13},
number = {7},
pages = {759-774},
doi = {https://doi.org/10.1002/env.527},
year = {2002}
}

@article{Chen2021,
  author    = {Chen, Y. and Koch, T. and Lim, K. G. and Xu, X. and Zakiyeva, N.},
  title     = {A review study of functional autoregressive models with application to energy forecasting},
  journal   = {WIREs Computational Statistics},
  volume    = {13},
  number    = {3},
  pages     = {e1525},
  year      = {2021},
  doi       = {10.1002/wics.1525}
}

@article{DeLivera2011,
author = {Alysha M. De Livera and Rob J. Hyndman and Ralph D. Snyder and},
title = {Forecasting Time Series With Complex Seasonal Patterns Using Exponential Smoothing},
journal = {Journal of the American Statistical Association},
volume = {106},
number = {496},
pages = {1513--1527},
year = {2011},
publisher = {ASA Website},
doi = {10.1198/jasa.2011.tm09771}
}

@ARTICLE{Lora2007,
  author={Lora, Alicia Troncoso and Santos, Jesus M. Riquelme and Exposito, Antonio Gomez and Ramos, Jose Luis MartÍnez and Santos, Jose C. Riquelme},
  journal={IEEE Transactions on Power Systems}, 
  title={Electricity Market Price Forecasting Based on Weighted Nearest Neighbors Techniques}, 
  year={2007},
  volume={22},
  number={3},
  pages={1294-1301},
  doi={10.1109/TPWRS.2007.901670}}

@article{Tajmouati2024,
author = {Tajmouati, Samya and Wahbi, Bouazza E. L. and Bedoui, Adel and Abarda, Abdallah and Dakkon, Mohamed},
title = {Applying k-nearest neighbors to time series forecasting: Two new approaches},
journal = {Journal of Forecasting},
volume = {43},
number = {5},
pages = {1559-1574},
doi = {https://doi.org/10.1002/for.3093},
url = {https://onlinelibrary.wiley.com/doi/abs/10.1002/for.3093},
eprint = {https://onlinelibrary.wiley.com/doi/pdf/10.1002/for.3093},
year = {2024}
}

@article{HNIN2024,
title = {A hybrid K-means and KNN approach for enhanced short-term load forecasting incorporating holiday effects},
journal = {Energy Reports},
volume = {12},
pages = {5942-5959},
year = {2024},
issn = {2352-4847},
doi = {https://doi.org/10.1016/j.egyr.2024.11.050},
url = {https://www.sciencedirect.com/science/article/pii/S235248472400773X},
author = {Su Wutyi Hnin and Jessada Karnjana and Youji Kohda and Chawalit Jeenanunta}
}

@Manual{Hyndman2025,
    title = {{forecast}: Forecasting functions for time series and
      linear models},
    author = {Rob Hyndman and George Athanasopoulos and Christoph
      Bergmeir and Gabriel Caceres and Leanne Chhay and Mitchell
      O'Hara-Wild and Fotios Petropoulos and Slava Razbash and Earo
      Wang and Farah Yasmeen},
    year = {2025},
    note = {R package version 8.24.0},
    url = {https://pkg.robjhyndman.com/forecast/}
}

@Article{Hyndman2008,
    title = {Automatic time series forecasting: the forecast package
      for {R}},
    author = {Rob J Hyndman and Yeasmin Khandakar},
    journal = {Journal of Statistical Software},
    volume = {27},
    number = {3},
    pages = {1--22},
    year = {2008},
    doi = {10.18637/jss.v027.i03}
}

@article{stanislaus2024,
  title={A comparative analysis of five time series models for CO2 emissions forecasting un Port-Harcourt and its environs},
  author={Stanislaus, Engr Orimadike Okechukwu and Adaku, Okengwu Ugochi and Odianonsen, Omijeh Bourdillon},
  journal={European Journal of Engineering and Technology},
  volume={12},
  number={1},
  year={2024}
}

@article{Tawiah2023,
author = {Tawiah, Kassim and Daniyal, Muhammad and Qureshi, Moiz},
title = {Pakistan CO2 Emission Modelling and Forecasting: A Linear and Nonlinear Time Series Approach},
journal = {Journal of Environmental and Public Health},
volume = {2023},
number = {1},
pages = {5903362},
doi = {https://doi.org/10.1155/2023/5903362},
url = {https://onlinelibrary.wiley.com/doi/abs/10.1155/2023/5903362},
eprint = {https://onlinelibrary.wiley.com/doi/pdf/10.1155/2023/5903362},
year = {2023}
}

@Article{su17041471,
AUTHOR = {Tian, Yaxin and Ren, Xiang and Li, Keke and Li, Xiangqian},
TITLE = {Carbon Dioxide Emission Forecast: A Review of Existing Models and Future Challenges},
JOURNAL = {Sustainability},
VOLUME = {17},
YEAR = {2025},
NUMBER = {4},
ARTICLE-NUMBER = {1471},
URL = {https://www.mdpi.com/2071-1050/17/4/1471},
ISSN = {2071-1050},
DOI = {10.3390/su17041471}
}

@article{Hastie1986,
author = {Trevor Hastie and Robert Tibshirani},
title = {{Generalized Additive Models}},
volume = {1},
journal = {Statistical Science},
number = {3},
publisher = {Institute of Mathematical Statistics},
pages = {297 -- 310},
year = {1986},
doi = {10.1214/ss/1177013604},
URL = {https://doi.org/10.1214/ss/1177013604}
}

@article{hastie2015package,
  title={Package 'gam'},
  author={Hastie, Trevor},
  journal={R package version},
  volume={90124},
  year={2015}
}

@book{ramsay2006,
  title={Functional Data Analysis},
  author={Ramsay, J. and Silverman, B.W.},
  isbn={9780387227511},
  lccn={2005923773},
  series={Springer Series in Statistics},
  url={https://books.google.it/books?id=REzuyz_V6OQC},
  year={2006},
  publisher={Springer New York}
}

@book{bosq2000linear,
  title={Linear processes in function spaces: theory and applications},
  author={Bosq, Denis},
  volume={149},
  year={2000},
  publisher={Springer Science \& Business Media}
}

@misc{Damon2024,
  author       = {Julien Damon and Serge Guillas},
  title        = {far: Modelization for Functional AutoRegressive Processes},
  year         = {2024},
  url          = {https://cran.r-project.org/web/packages/far/index.html},
  doi          = {10.32614/CRAN.package.far},
  note         = {R package version 0.6-7},
  publisher    = {CRAN}
}

@book{surowiecki2005,
  title={The wisdom of crowds},
  author={Surowiecki, James},
  year={2005},
  publisher={Vintage}
}

@article{CLEMEN1989,
title = {Combining forecasts: A review and annotated bibliography},
journal = {International Journal of Forecasting},
volume = {5},
number = {4},
pages = {559-583},
year = {1989},
issn = {0169-2070},
doi = {https://doi.org/10.1016/0169-2070(89)90012-5},
url = {https://www.sciencedirect.com/science/article/pii/0169207089900125},
author = {Robert T. Clemen}
}

@incollection{TIMMERMANN2006,
title = {Chapter 4 Forecast Combinations},
booktitle = {Handbook of Economic Forecasting},
editor = {G. Elliott and C.W.J. Granger and A. Timmermann},
series = {Handbook of Economic Forecasting},
publisher = {Elsevier},
volume = {1},
pages = {135-196},
year = {2006},
issn = {1574-0706},
doi = {https://doi.org/10.1016/S1574-0706(05)01004-9},
url = {https://www.sciencedirect.com/science/article/pii/S1574070605010049},
author = {Allan Timmermann}
}

@article{WANG2023,
title = {Forecast combinations: An over 50-year review},
journal = {International Journal of Forecasting},
volume = {39},
number = {4},
pages = {1518-1547},
year = {2023},
issn = {0169-2070},
doi = {https://doi.org/10.1016/j.ijforecast.2022.11.005},
url = {https://www.sciencedirect.com/science/article/pii/S0169207022001480},
author = {Xiaoqian Wang and Rob J. Hyndman and Feng Li and Yanfei Kang}
}

@article{BUNN1985151,
title = {Statistical efficiency in the linear combination of forecasts},
journal = {International Journal of Forecasting},
volume = {1},
number = {2},
pages = {151-163},
year = {1985},
issn = {0169-2070},
doi = {https://doi.org/10.1016/0169-2070(85)90020-2},
url = {https://www.sciencedirect.com/science/article/pii/0169207085900202},
author = {Derek W. Bunn}
}

@article{GENRE2013108,
title = {Combining expert forecasts: Can anything beat the simple average?},
journal = {International Journal of Forecasting},
volume = {29},
number = {1},
pages = {108-121},
year = {2013},
issn = {0169-2070},
doi = {https://doi.org/10.1016/j.ijforecast.2012.06.004},
url = {https://www.sciencedirect.com/science/article/pii/S016920701200088X},
author = {Véronique Genre and Geoff Kenny and Aidan Meyler and Allan Timmermann}
}

@article{Diebold1995,
author = {Francis X. Diebold and Roberto S. Mariano},
title = {Comparing Predictive Accuracy},
journal = {Journal of Business \& Economic Statistics},
volume = {13},
number = {3},
pages = {253--263},
year = {1995},
publisher = {ASA Website},
doi = {10.1080/07350015.1995.10524599}
}

@article{Wickramasuriya2019,
author = {Shanika L. Wickramasuriya and George Athanasopoulos and Rob J. Hyndman and},
title = {Optimal Forecast Reconciliation for Hierarchical and Grouped Time Series Through Trace Minimization},
journal = {Journal of the American Statistical Association},
volume = {114},
number = {526},
pages = {804--819},
year = {2019},
publisher = {ASA Website},
doi = {10.1080/01621459.2018.1448825},
URL = {https://doi.org/10.1080/01621459.2018.1448825},
eprint = {https://doi.org/10.1080/01621459.2018.1448825}
}

@article{GIROLIMETTO2024a,
title = {Cross-temporal probabilistic forecast reconciliation: Methodological and practical issues},
journal = {International Journal of Forecasting},
volume = {40},
number = {3},
pages = {1134-1151},
year = {2024},
issn = {0169-2070},
doi = {https://doi.org/10.1016/j.ijforecast.2023.10.003},
url = {https://www.sciencedirect.com/science/article/pii/S0169207023001024},
author = {Daniele Girolimetto and George Athanasopoulos and Tommaso {Di Fonzo} and Rob J. Hyndman}
}

@article{DIFONZO2024,
title = {Forecast combination-based forecast reconciliation: Insights and extensions},
journal = {International Journal of Forecasting},
volume = {40},
number = {2},
pages = {490-514},
year = {2024},
issn = {0169-2070},
doi = {https://doi.org/10.1016/j.ijforecast.2022.07.001},
url = {https://www.sciencedirect.com/science/article/pii/S0169207022000991},
author = {Tommaso {Di Fonzo} and Daniele Girolimetto}
}

@misc{girolimetto2024b,
      title={Coherent forecast combination for linearly constrained multiple time series}, 
      author={Girolimetto, D. and T. {Di Fonzo}},
      year={2024},
      note={2412.03429},
      archivePrefix={arXiv},
      primaryClass={stat.ME},
      url={https://arxiv.org/abs/2412.03429}
}

@article{Sapio2016,
author = {Sapio, A. and Spagnolo, N.},
title = {Price regimes in an energy island: Tacit collusion vs. cost and network explanations},
journal = {Energy Economics},
volume = {55},
number = {},
pages = {157-172},
year = {2016},
}

@article{Sapio2020,
author = {Sapio, A. and Spagnolo, N.},
title = {The effect of a new power cable on energy prices volatility spillovers.},
journal = {Energy Policy},
volume = {144},
number = {},
pages = {111488},
year = {2020},
}

@article{Weinhold2021,
author = {Weinhold, R. and Mieth, R.},
title = {Power Market Tool (POMATO) for the analysis of zonal electricity markets},
journal = {SoftwareX},
volume = {16},
number = {},
pages = {100870},
year = {2021},
}

@article{Hansen2011,
 ISSN = {00129682, 14680262},
 URL = {http://www.jstor.org/stable/41057463},
 author = {Peter R. Hansen and Asger Lunde and James M. Nason},
 journal = {Econometrica},
 number = {2},
 pages = {453--497},
 publisher = {[Wiley, Econometric Society]},
 title = {THE MODEL CONFIDENCE SET},
 urldate = {2025-12-01},
 volume = {79},
 year = {2011}
}

@misc{bernardi2014,
      title={{The Model Confidence Set package for R}}, 
      author={Mauro Bernardi and Leopoldo Catania},
      year={2014},
      eprint={1410.8504},
      archivePrefix={arXiv},
      primaryClass={stat.CO},
      url={https://arxiv.org/abs/1410.8504}, 
}

\paragraph{Declaration of interests} The authors report that financial support was provided by European Union - NextGenerationEU, Mission 4, Component 2, in the framework of the GRINS - Growing Resilient, INclusive and Sustainable project (GRINS PE00000018 - CUP C93C22005270001).

\paragraph{Acknowledgments} This study was funded by the European Union - NextGenerationEU, Mission 4, Component 2, in the framework of the GRINS - Growing Resilient, INclusive and Sustainable project (GRINS PE00000018 - CUP C93C22005270001). The views and opinions expressed are solely those of the authors and do not necessarily reflect those of the European Union, nor can the European Union be held responsible for them.

\paragraph{Data availability statement} Data are available from the authors. 

\appendix
\renewcommand{\thefigure}{A\arabic{figure}} % Equation numbering: A1, A2, 
\setcounter{figure}{0}  % Reset equation counter at start of appendix

\begin{landscape}

\begin{figure}[H]
\section{Figures}\label{app:Figures}
\centering
    \includegraphics[width=0.8\linewidth]{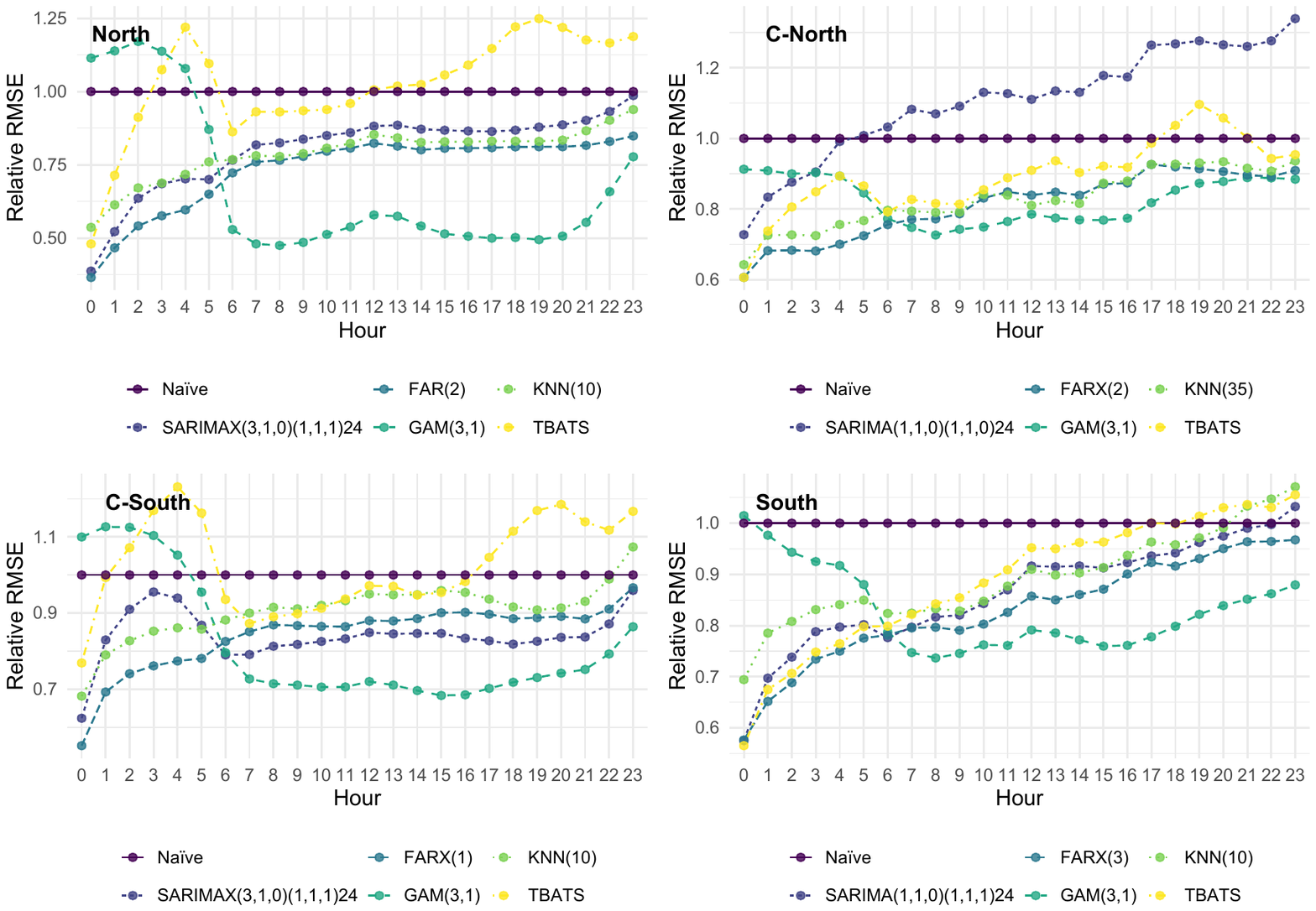}
    \caption{Relative RMSE of single models for North, Centre-North, Centre-South, and South, compared to the Na\"ive model. Test set.}
    \label{fig:test_singm_marketzones}
\end{figure}

\begin{figure}[H]
    \centering
    \includegraphics[width=0.8\linewidth]{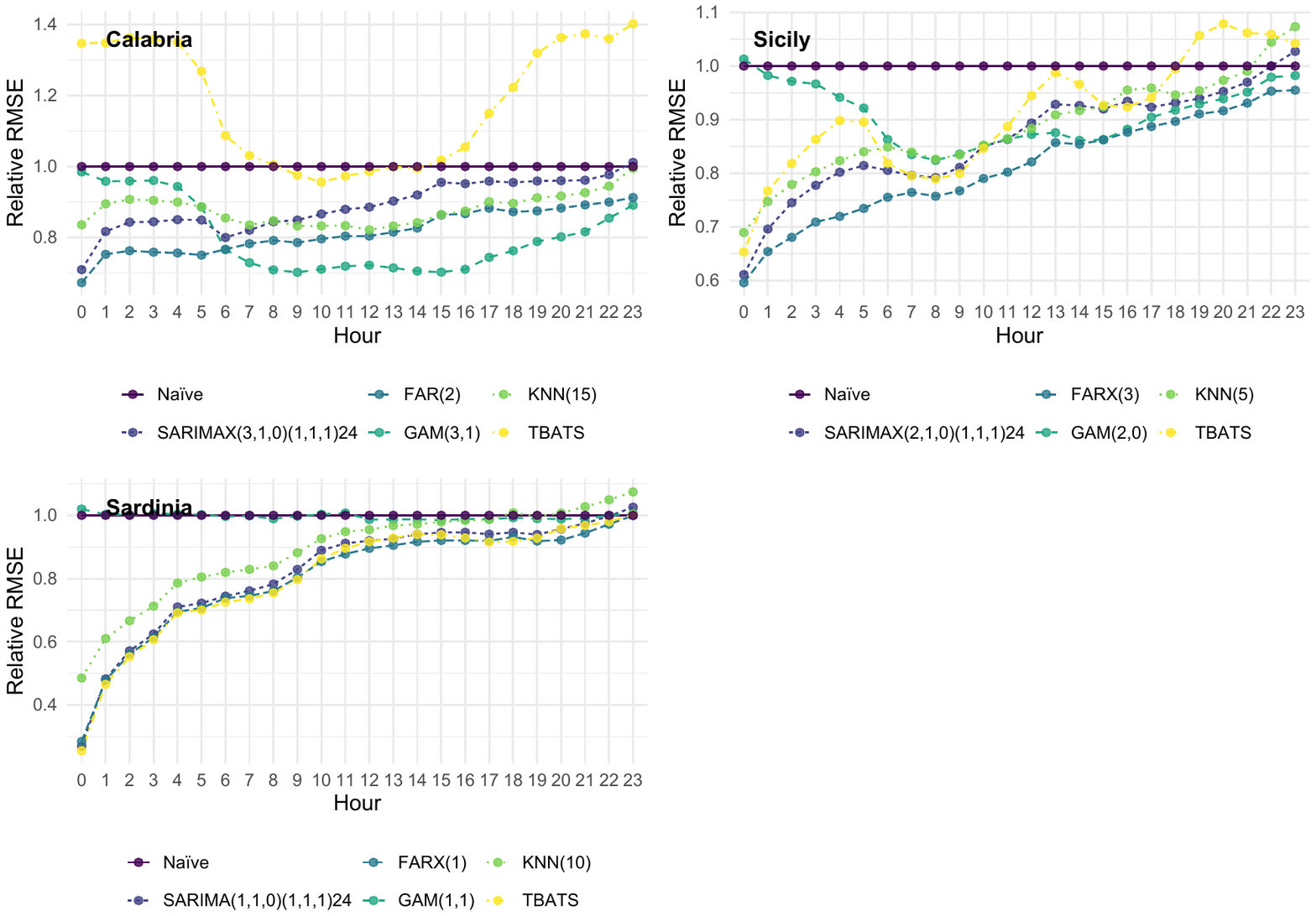}
    \caption{Relative RMSE of single models for Calabria, Sicily and Sardinia, compared to the Na\"ive model. Test set.}
    \label{fig:test_singm_marketzones2}
\end{figure}
\end{landscape}

\renewcommand{\thetable}{B\arabic{table}} % Equation numbering: A1, A2, ...
\setcounter{table}{0}  % Reset equation counter at start of appendix

\section{Tables}\label{app:Tables}

\begin{table}[H]
\caption{Summary statistics of carbon emissions by Italian electricity market zone, reported for the training, calibration, and test sets.}
\label{tab:descriptive_stat}
\centering
\resizebox{10cm}{!}{
\begin{tabular}{lccccc}
  \hline
  Zone & Mean & Median & Stdev & Skewness & Kurtosis \\ 
  \hline
  \multicolumn{5}{l}{\textit{Training set (January 2021--June 2022)}} \\
   Italy & 13425.65 & 13126.30 & 4284.71 & 0.29 & -0.57 \\ 
   North & 6332.16 & 6116.99 & 2575.86 & 0.37 & -0.52 \\ 
   Cnorth & 480.22 & 471.12 & 160.53 & -0.42 & -0.60 \\ 
   Csouth & 1652.04 & 1470.48 & 873.56 & 0.55 & -0.59 \\ 
   South & 1758.90 & 1711.79 & 652.22 & 0.28 & -0.54 \\ 
   Calabria & 886.40 & 859.62 & 503.68 & 0.23 & -0.91 \\ 
   Sicily & 863.01 & 819.00 & 375.55 & 0.53 & -0.16 \\ 
   Sardinia & 1452.91 & 1451.50 & 248.65 & -0.07 & 0.34 \\ 
  \hline
  \multicolumn{5}{l}{\textit{Calibration set (July 2022--December 2022)}} \vspace{0.1cm}\\
 Italy &  14985.72 & 14910.04 & 4076.00 & 0.05 & -0.81 \\ 
 North  & 6379.07 & 6416.48 & 2389.50 & 0.13 & -0.63 \\ 
 Cnorth  & 519.92 & 542.57 & 156.71 & -0.63 & -0.05 \\ 
 Csouth  & 1665.10 & 1511.20 & 959.33 & 0.59 & -0.37 \\ 
 South  & 2279.28 & 2224.49 & 579.57 & 0.43 & -0.04 \\ 
 Calabria  & 763.24 & 735.33 & 511.69 & 0.11 & -1.09 \\ 
 Sicily  & 1632.44 & 1667.61 & 321.88 & -0.51 & 0.40 \\ 
 Sardinia  & 1746.66 & 1760.30 & 190.09 & -0.67 & 1.39 \\ 
   \hline
  \multicolumn{5}{l}{\textit{Test set (January 2023--December 2023)}} \vspace{0.1cm}\\
 Italy  & 12924.16 & 12560.81 & 4066.78 & 0.38 & -0.45 \\ 
 North  & 6114.24 & 5874.48 & 2495.23 & 0.43 & -0.37 \\ 
 Cnorth  & 486.24 & 516.52 & 164.99 & -0.57 & -0.62 \\ 
 Csouth & 1609.28 & 1403.43 & 844.97 & 0.66 & -0.46 \\ 
 South & 1623.33 & 1576.45 & 601.40 & 0.40 & -0.38 \\ 
 Calabria & 864.85 & 817.20 & 479.09 & 0.34 & -0.93 \\ 
 Sicily & 791.37 & 719.36 & 357.30 & 0.70 & -0.11 \\ 
 Sardinia  & 1434.85 & 1447.22 & 228.37 & -0.52 & 0.43 \\ 
   \hline
\end{tabular}
}
\end{table}

\begin{table}[H]
\caption{Hourly RMSE, hourly average RMSE and R$^2$ in the test set for Italy. The best hourly average performance in bold.}
\label{tab:test_sing_models_italy}
\centering
\resizebox{15cm}{!}{
\begin{tabular}{lcccccc}
\hline
Hour  & Na\"ive  & SARIMAX(2,1,0)(1,1,1)24 & FARX(2) & GAM(3,1)  & TBATS   & KNN(10)   \\ \hline
0     & 1747.82 & 603.62                  & 586.27  & 2014.15 & 828.55  & 895.14  \\
1     & 1649.77 & 772.52                  & 724.59  & 1913.53 & 1076.61 & 968.44  \\
2     & 1576.46 & 887.19                 & 801.13  & 1835.65 & 1247.85 & 1010.81 \\
3     & 1579.05 & 979.06                 & 865.96  & 1787.27 & 1555.41 & 1050.16 \\
4     & 1601.69 & 1039.11                 & 930.96  & 1714.19 & 1969.83 & 1101.75 \\
5     & 1801.48 & 1197.70                 & 1152.19 & 1572.13 & 2204.72 & 1280.53 \\
6     & 2634.33 & 1985.94                 & 1927.47 & 1448.36 & 2343.35 & 1899.96 \\
7     & 3408.24 & 2739.02                 & 2627.1  & 1685.3  & 2999.86 & 2488.23 \\
8     & 3643.49 & 2964.60                  & 2829.13 & 1777.15 & 3255.62 & 2671.11 \\
9     & 3683.27 & 3026.55                 & 2892.96 & 1850.17 & 3358.22 & 2722.47 \\
10    & 3701.08 & 3079.08                 & 2949.4  & 1941.42 & 3457.43 & 2776.08 \\
11    & 3650.4  & 3065.49                 & 2944.01 & 1985.57 & 3471.27 & 2793.27 \\
12    & 3328.83 & 2830.48                 & 2727.41 & 1926.09 & 3280.71 & 2616.79 \\
13    & 3398.8  & 2888.15                 & 2759.18 & 1954.43 & 3350.12 & 2641.73 \\
14    & 3650.12 & 3075.39                 & 2933.18 & 2017    & 3641.52 & 2812.86 \\
15    & 3745.77 & 3174.88                 & 3059.69 & 2005.06 & 3911.17 & 2950.08 \\
16    & 3725.02 & 3158.19                 & 3039.63 & 1978.08 & 4046.44 & 2975.35 \\
17    & 3526.17 & 2995.49                 & 2882.34 & 1883.04 & 4003.37 & 2884.23 \\
18    & 3311.3  & 2803.59                  & 2704.5  & 1791.15 & 3875.57 & 2708.54 \\
19    & 3170.96 & 2717.13                 & 2615.32 & 1715.96 & 3781.29 & 2614.37 \\
20    & 3033.02 & 2628.65                 & 2514.62 & 1682.37 & 3533.3  & 2533.37 \\
21    & 2683.83 & 2376.60                 & 2246.8  & 1619.47 & 2978.78 & 2337.93 \\
22    & 2330.99 & 2136.75                 & 1979.21 & 1613.7  & 2497.61 & 2150.83 \\
23    & 2017.95 & 1956.45                 & 1774.22 & 1613.48 & 2166.09 & 1988.89 \\
\hline
Ave.  & 2858.33 & 2295.07                 & 2186.14 & \textbf{1805.2}  & 2868.11 & 2203.04 \\
R$^2$ & 0.49    & 0.65                    & 0.68    & \textbf{0.81}    & 0.47    & 0.69    \\
\hline
\end{tabular}
}
\end{table}

\begin{table}[H]
\caption{Hourly RMSE, hourly average RMSE and R$^2$ in the calibration set for Italy. The best hourly average performance in bold.}
\label{tab:calb_singm_italy}
\centering
\resizebox{14cm}{!}{
\begin{tabular}{lcccccc}
\hline
Hour                                       & Na\"ive   & SARIMAX(2,1,0)(1,1,1)24 & FARX(2)          & GAM(3,1)       & TBATS   & KNN(10)   \\ \hline
0                    & 1934.4  & 783.63                  & 726.32  & 1901.91          & 1165.37 & 1047.4  \\
1                    & 1850.62 & 999.65                  & 852.96  & 1839.2           & 1547.37 & 1131.61 \\
2                    & 1787.36 & 1082.1                  & 893.17  & 1744.02          & 1803.55 & 1130.84 \\
3                    & 1732.41 & 1185.39                 & 939.54  & 1657.98          & 2201.58 & 1161.25 \\
4                    & 1730.57 & 1234.81                 & 979.79  & 1609.97          & 2599.61 & 1170.56 \\
5                    & 1993.67 & 1435.93                 & 1219.38 & 1488.41          & 2738.53 & 1384.42 \\
6                    & 3019.32 & 2299.66                 & 2032.11          & 1495.9  & 2792.9  & 2139.71 \\
7                    & 3802.94 & 2994.56                 & 2623.42          & 1816.09 & 3302.3  & 2664.07 \\
8                    & 4006.97 & 3148.54                 & 2768.9           & 1944.74 & 3461.96 & 2759.48 \\
9                    & 4054.56 & 3183.03                 & 2810.67          & 1984.83 & 3510.49 & 2775.11 \\
10                   & 4032.99 & 3177.17                 & 2834.07          & 2011.23 & 3544.32 & 2792.73 \\
11                   & 3984.19 & 3154.29                 & 2840.02          & 2022.3  & 3554.66 & 2779.49 \\
12                   & 3859.39 & 3094.95                 & 2810.18          & 2006.73 & 3526.03 & 2745.07 \\
13                   & 3912.7  & 3164.33                 & 2857.08          & 2047.56 & 3666.78 & 2832.36 \\
14                   & 4081.13 & 3297.02                 & 2954.62          & 2106.45 & 3991.52 & 2966.27 \\
15                   & 4137.96 & 3334.79                 & 2992.64          & 2143.02 & 4191.56 & 3000.32 \\
16                   & 4059.9  & 3317.88                 & 2975.58          & 2100.58 & 4289.79 & 2967.97 \\
17                   & 3761.57 & 3145.27                 & 2777.73          & 1946.52 & 4145.6  & 2804.65 \\
18                   & 3451.93 & 2966.86                 & 2584.59          & 1809.17 & 3918.66 & 2628.55 \\
19                   & 3282.8  & 2921.25                 & 2487.29          & 1722.14 & 3762.69 & 2539.81 \\
20                   & 3212.19 & 2887.36                 & 2416.48          & 1703.04 & 3590.9  & 2525.25 \\
21                   & 2987.37 & 2772.98                 & 2278.99          & 1701.77 & 3272.09 & 2443.42 \\
22                   & 2642.74 & 2531.22                 & 2055.81          & 1731.99 & 2854.17 & 2265.98 \\
23                   & 2315.08 & 2333.09                 & 1901.82          & 1669.91 & 2541.37 & 2114.49 \\
\hline
Ave.  & 3151.45 & 2518.57                 & 2192.22          & \textbf{1841.89} & 3165.58 & 2282.12   \\
R$^2$ & 0.36    & 0.58                    & 0.67             & \textbf{0.79}    & 0.36    & 0.66   \\ \hline
\end{tabular}
}
\end{table}

\begin{landscape}
\begin{table}[H]
\caption{Hourly RMSE, hourly average RMSE and R$^2$ in the test set for Italy. The best hourly average performance in bold.}
\label{tab:test_fcomb_italy}
\centering
\resizebox{23cm}{!}{
\begin{tabular}{lcccccccccccc}
\hline
Hour  &  Na\"ive  & COMB1.SA & COMB1.BG & COMB2.SA & COMB2.BG & COMB2.SEL & COMB3.SA & COMB3.BG & COMB3.SEL & COMB4.SA & COMB4.BG & COMB4.SEL \\ \hline
0     & 1747.82 & 661.87   & 604.23   & 1097.32  & 1289.24  & 586.27    & 1060.16  & 582.88   & 603.62    & 1217.07  & 1298.32  & 895.14    \\
1     & 1649.77 & 722.76   & 766.5    & 1056.15  & 1228.41  & 724.59    & 1012.8   & 721.35   & 772.52    & 1173.12  & 1244.59  & 968.44    \\
2     & 1576.46 & 751.29   & 869.63   & 1024.44  & 1181.9   & 801.13    & 970.41   & 815.59   & 887.19    & 1130.9   & 1195.64  & 1010.81   \\
3     & 1579.05 & 799.52   & 992.32   & 1022     & 1165.87  & 865.96    & 955.49   & 896.18   & 979.06    & 1111.51  & 1170.88  & 1050.16   \\
4     & 1601.69 & 864.62   & 1127.82  & 992.32   & 1121.55  & 930.96    & 921.9    & 947.2    & 1039.11   & 1087.43  & 1138.49  & 1101.75   \\
5     & 1801.48 & 1011.65  & 1303.43  & 964.88   & 1050.86  & 1152.19   & 907.31   & 1094.69  & 1197.7    & 1082.93  & 1110.77  & 1280.53   \\
6     & 2634.33 & 1601.77  & 1941.65  & 1310.53  & 1263.71  & 1448.36   & 1351.69  & 1863.71  & 1448.36   & 1391.2   & 1363.55  & 1448.36   \\
7     & 3408.24 & 2237.78  & 2654.42  & 1842.86  & 1737.8   & 1685.3    & 1941.4   & 2600.42  & 1685.3    & 1861.02  & 1811.36  & 1685.3    \\
8     & 3643.49 & 2432.35  & 2873.78  & 2011.24  & 1893.5   & 1777.15   & 2119.54  & 2820.3   & 1777.15   & 2010.65  & 1955.59  & 1777.15   \\
9     & 3683.27 & 2496.06  & 2939.24  & 2080.83  & 1964.83  & 1850.17   & 2184.94  & 2882.21  & 1850.17   & 2064.47  & 2010.88  & 1850.17   \\
10    & 3701.08 & 2555.58  & 2997.02  & 2150.89  & 2039.8   & 1941.42   & 2244.29  & 2934.31  & 1941.42   & 2123.47  & 2072.37  & 1941.42   \\
11    & 3650.4  & 2555.23  & 2988.63  & 2167.88  & 2063.17  & 1985.57   & 2243.89  & 2921.13  & 1985.57   & 2145.5   & 2096.17  & 1985.57   \\
12    & 3328.83 & 2354.56  & 2763.59  & 2019.86  & 1934.23  & 1926.09   & 2073.19  & 2693.34  & 1926.09   & 2012.98  & 1970.96  & 1926.09   \\
13    & 3398.8  & 2392.11  & 2818.71  & 2054.03  & 1968.13  & 1954.43   & 2120.25  & 2749.74  & 1954.43   & 2046.4   & 2004.71  & 1954.43   \\
14    & 3650.12 & 2558.27  & 3019.46  & 2179.45  & 2080.13  & 2017      & 2261.93  & 2931.77  & 2017      & 2178.56  & 2130.14  & 2017      \\
15    & 3745.77 & 2670.54  & 3152.49  & 2242.63  & 2126.29  & 2005.06   & 2316.39  & 3025.95  & 2005.06   & 2248.93  & 2191.02  & 2005.06   \\
16    & 3725.02 & 2680.88  & 3174.75  & 2224.87  & 2107.52  & 1978.08   & 2308.07  & 3011.56  & 1978.08   & 2260.25  & 2199.11  & 1978.08   \\
17    & 3526.17 & 2566.7   & 3051.27  & 2107.76  & 1997.41  & 1883.04   & 2187.39  & 2855.74  & 1883.04   & 2172.6   & 2111.04  & 1883.04   \\
18    & 3311.3  & 2416.13  & 2885.09  & 1984.75  & 1884.32  & 1791.15   & 2047.19  & 2671.57  & 1791.15   & 2041.31  & 1984.86  & 1791.15   \\
19    & 3170.96 & 2326.35  & 2800.29  & 1902.62  & 1803.4   & 1715.96   & 1959.16  & 2584.74  & 1715.96   & 1943.27  & 1887.62  & 1715.96   \\
20    & 3033.02 & 2231.57  & 2682.26  & 1835.42  & 1744.21  & 1682.37   & 1890.89  & 2498.68  & 1682.37   & 1877.43  & 1824.59  & 1682.37   \\
21    & 2683.83 & 1971.42  & 2374.18  & 1649.94  & 1583.96  & 1619.47   & 1706.85  & 2254.15  & 1619.47   & 1723.54  & 1678.8   & 1619.47   \\
22    & 2330.99 & 1745.92  & 2090.49  & 1502.92  & 1470.61  & 1613.7    & 1555.33  & 2023.42  & 1613.7    & 1597.46  & 1564.53  & 1613.7    \\
23    & 2017.95 & 1587.52  & 1876.65  & 1403.28  & 1397.58  & 1613.48   & 1463.18  & 1853.92  & 1613.48   & 1503.73  & 1481.82  & 1613.48   \\
\hline
Ave.  & 2858.33 & 1924.69  & 2281.16  & 1701.2   & 1670.77  & \textbf{1564.54}   & 1741.82  & 2176.44  & 1581.96   & 1750.24  & 1729.08  & 1616.44   \\
R$^2$ & 0.49    & 0.76     & 0.66     & 0.82     & 0.83     & \textbf{0.85}      & 0.81     & 0.69     & 0.85      & 0.81     & 0.82     & 0.84      \\ \hline
\end{tabular}
}
\end{table}
\end{landscape}

\begin{table}[H]
    \caption{Diebold-Mariano test p-values for pairwise comparisons in the North. H$_0$ equal forecast accuracy; H$_1$: the column model outperforms the row model.}
    \label{tab.DMtestnorth}
    \centering
    \resizebox{15.5cm}{!}{
\begin{tabular}{lcccccccc}
  \hline
 & SARIMAX$^a$ & FAR(2) & GAM(3,1) & TBATS & KNN(10) & COMB2.SEL & COMB3.SEL \\ 
  \hline
SARIMAX$^a$ & - & $<0.05$ & $<0.05$ & 1.00 & $<0.05$ & $<0.05$ & $<0.05$ \\ 
  FAR(2) & 1.00 & - & $<0.05$ & 1.00 & 1.00 & $<0.05$ & $<0.05$\\ 
  GAM(3,1) & 1.00 & 1.00 & - & 1.00 & 1.00 & $<0.05$ & $<0.05$  \\ 
  TBATS & $<0.05$ & $<0.05$ & $<0.05$ & - & $<0.05$ & $<0.05$ & $<0.05$  \\ 
  KNN(10) & 1.00 & $<0.05$ & $<0.05$ & 1.00 & - & $<0.05$ & $<0.05$  \\ 
  COMB2.SEL & 1.00 & 1.00 & 1.00 & 1.00 & 1.00 & - & 1.00  \\ 
  COMB3.SEL & 1.00 & 1.00 & 1.00 & 1.00 & 1.00 & $<0.05$ & -  \\ 
  
   \hline
\end{tabular}
    }
\begin{tablenotes}
\item[]{\footnotesize \textit{Note}. $^a$ the model order is (3,1,0)(1,1,1)24.}
\end{tablenotes}
\end{table}

\begin{table}[H]
    \caption{Diebold-Mariano test p-values for pairwise comparisons for Centre-North.}
    \label{tab.DMtestcnorth}
    \centering
    \resizebox{15.5cm}{!}{
\begin{tabular}{lccccccc}
  \hline
 & SARIMA$^a$ & FARX(2) & GAM(3,1) & TBATS & KNN(35) & COMB2.SA & COMB2.SEL \\ 
  \hline
SARIMA$^a$ & - & $<0.05$ & $<0.05$ & $<0.05$ & $<0.05$ & $<0.05$ & $<0.05$ \\ 
  FARX(2) & 1.00 & - & 0.15 & 1.00 & 1.00 & $<0.05$ & $<0.05$ \\ 
  GAM(3,1) & 1.00 & 0.85 & - & 1.00 & 1.00 & $<0.05$ & $<0.05$ \\ 
  TBATS & 1.00 & $<0.05$ & $<0.05$ &  - & $<0.05$ & $<0.05$ & $<0.05$ \\ 
  KNN(35) & 1.00 & $<0.05$ & $<0.05$ & 1.00 & - & $<0.05$ & $<0.05$ \\ 
  COMB2.SA & 1.00 & 1.00 & 1.00 & 1.00 & 1.00 & - & 0.88 \\ 
  COMB2.SEL & 1.00 & 1.00 & 1.00 & 1.00 & 1.00 & 0.12 & - \\
   \hline
\end{tabular}
}
    \begin{tablenotes}
\item[]{\footnotesize \textit{Note}. $^a$ the model order is (1,1,0)(1,1,0)24.}
\end{tablenotes}
\end{table}

\begin{table}[H]
    \caption{Diebold-Mariano test p-values for pairwise comparisons for Centre-South.}
    \label{tab.DMtestcsouth}
    \centering
    \resizebox{15.5cm}{!}{
\begin{tabular}{lccccccc}
  \hline
 & SARIMAX$^a$ & FARX(1) & GAM(3,1) & TBATS & KNN(10) & COMB3.SA & COMB2.SEL \\ 
  \hline
SARIMAX$^a$ & - & 1.00 & $<0.05$ & 1.00 & 1.00 & $<0.05$ & $<0.05$ \\ 
  FARX(1) & $<0.05$ & - & $<0.05$ & 1.00 & 1.00 & $<0.05$ & $<0.05$ \\ 
  GAM(3,1) & 1.00 & 1.00 & - & 1.00 & 1.00 & $<0.05$ & $<0.05$ \\ 
  TBATS & $<0.05$ & $<0.05$ & $<0.05$ & - & $<0.05$ & $<0.05$ & $<0.05$ \\ 
  KNN(10) & $<0.05$ & $<0.05$ & $<0.05$ & 1.00 & - & $<0.05$ & $<0.05$ \\ 
  COMB3.SA & 1.00 & 1.00 & 1.00 & 1.00 & 1.00 & - & 1.00 \\ 
  COMB2.SEL & 1.00 & 1.00 & 1.00 & 1.00 & 1.00 & $<0.05$ & - \\ 
   \hline
\end{tabular}
    }
    \begin{tablenotes}
\item[]{\footnotesize \textit{Note}. $^a$ the model order is (3,1,0)(1,1,1)24.}
\end{tablenotes}
\end{table}

\begin{table}[H]
    \caption{Diebold-Mariano test p-values for pairwise comparisons for the South.}
    \label{tab.DMtestsouth}
    \centering
    \resizebox{15.5cm}{!}{
\begin{tabular}{lccccccc}
  \hline
 & SARIMA$^a$ & FARX(3) & GAM(3,1) & TBATS & KNN(10) & COMB3.SA & COMB2.SEL \\ 
  \hline
SARIMA$^a$ & - & $<0.05$ & $<0.05$ & 1.00 & 1.00 & $<0.05$ & $<0.05$ \\ 
  FARX(3) & 1.00 & - & $<0.05$ & 1.00 & 1.00 & $<0.05$ & $<0.05$ \\ 
  GAM(3,1) & 1.00 & 1.00 & - & 1.00 & 1.00 & $<0.05$ & $<0.05$ \\ 
  TBATS & $<0.05$ & $<0.05$ & $<0.05$ & - & $<0.05$ & $<0.05$ & $<0.05$ \\ 
  KNN(10) & $<0.05$ & $<0.05$ & $<0.05$ & 1.00 & - & $<0.05$ & $<0.05$ \\ 
  COMB3.SA & 1.00 & 1.00 & 1.00 & 1.00 & 1.00 & - & 1.00 \\ 
  COMB2.SEL & 1.00 & 1.00 & 1.00 & 1.00 & 1.00 & $<0.05$ &  - \\ 
   \hline
\end{tabular}
    }
\begin{tablenotes}
\item[]{\footnotesize \textit{Note}. $^a$ the model order is (1,1,0)(1,1,1)24.}
\end{tablenotes}
\end{table}

\begin{table}[H]
    \caption{Diebold-Mariano test p-values for pairwise comparisons for Calabria.}
    \label{tab.DMtestcalabria}
    \centering
    \resizebox{15.5cm}{!}{
\begin{tabular}{lccccccc}
  \hline
 & SARIMAX$^a$ & FAR(2) & GAM(3,1) & TBATS & KNN(15) & COMB3.SA & COMB2.SEL \\ 
  \hline
SARIMAX$^a$ & - & $<0.05$ & $<0.05$ & 1.00 & $<0.05$ & $<0.05$ & $<0.05$ \\ 
  FAR(2) & 1.00 & - & $<0.05$ & 1.00 & 1.00 & $<0.05$ & $<0.05$ \\ 
  GAM(3,1) & 1.00 & 1.00 & - & 1.00 & 1.00 & $<0.05$ & $<0.05$ \\ 
  TBATS & $<0.05$ & $<0.05$ & $<0.05$ & - & $<0.05$ & $<0.05$ & $<0.05$ \\ 
  KNN(15) & 1.00 & $<0.05$ & $<0.05$ & 1.00 & - & $<0.05$ & $<0.05$ \\ 
  COMB3.SA & 1.00 & 1.00 & 1.00 & 1.00 & 1.00 & - & 0.72 \\ 
  COMB2.SEL & 1.00 & 1.00 & 1.00 & 1.00 & 1.00 & 0.28 & - \\ 
   \hline
\end{tabular}
    }
    \begin{tablenotes}
\item[]{\footnotesize \textit{Note}. $^a$ the model order is (3,1,0)(1,1,1)24.}
\end{tablenotes}
\end{table}

\begin{table}[H]
    \caption{Diebold-Mariano test p-values for pairwise comparisons for Sicily.}
    \label{tab.DMtestsicily}
    \centering
    \resizebox{15.5cm}{!}{
\begin{tabular}{lccccccc}
  \hline
 & SARIMAX$^a$ & FARX(3) & GAM(2,0) & TBATS & KNN(5) & COMB3.SA & COMB2.BG \\ 
  \hline
SARIMAX$^a$ & - & $<0.05$ & 1.00 & 1.00 & 1.00 & $<0.05$ & $<0.05$ \\ 
  FARX(3) & 1.00 & - & 1.00 & 1.00 & 1.00 & $<0.05$ & $<0.05$ \\ 
  GAM(2,0) & $<0.05$ & $<0.05$ & - & 0.92 & 0.04 & $<0.05$ & $<0.05$ \\ 
  TBATS & $<0.05$ & $<0.05$ & 0.08 & - & $<0.05$ & $<0.05$ & $<0.05$ \\ 
  KNN(5) & $<0.05$ & $<0.05$ & 0.96 & 1.00 & - & $<0.05$ & $<0.05$ \\ 
  COMB3.SA & 1.00 & 1.00 & 1.00 & 1.00 & 1.00 & - & 1.00 \\ 
  COMB2.BG & 1.00 & 1.00 & 1.00 & 1.00 & 1.00 & $<0.05$ & - \\ 
   \hline
\end{tabular}
    }
        \begin{tablenotes}
\item[]{\footnotesize \textit{Note}. $^a$ the model order is (2,1,0)(1,1,1)24.}
\end{tablenotes}
\end{table}

\begin{table}[H]
    \caption{Diebold-Mariano test p-values for pairwise comparisons for Sardinia.}
    \label{tab.DMtestsardinia}
    \centering
    \resizebox{15.5cm}{!}{
\begin{tabular}{lccccccc}
  \hline
 & SARIMAX$^a$ & FARX(1) & GAM(1,1) & TBATS & KNN(10) & COMB1.BG & COMB1.SA \\ 
  \hline
SARIMAX$^a$ & - & $<0.05$ & 1.00 & $<0.05$ & 1.00 & $<0.05$ & $<0.05$ \\ 
  FARX(1) & 1.00 & - & 1.00 & 1.00 & 1.00 & $<0.05$ & $<0.05$ \\ 
  GAM(1,1) & $<0.05$ & $<0.05$ & - & $<0.05$ & $<0.05$ & $<0.05$ & $<0.05$ \\ 
  TBATS & 1.00 & $<0.05$ & 1.00 & - & 1.00 & $<0.05$ & $<0.05$ \\ 
  KNN(10) & $<0.05$ & $<0.05$ & 1.00 & $<0.05$ & - & $<0.05$ & $<0.05$ \\ 
  COMB1.BG & 1.00 & 1.00 & 1.00 & 1.00 & 1.00 & - & 1.00 \\ 
  COMB1.SA & 1.00 & 1.00 & 1.00 & 1.00 & 1.00 & $<0.05$ & - \\ 
   \hline
\end{tabular}
    }
            \begin{tablenotes}
\item[]{\footnotesize \textit{Note}. $^a$ the model order is (1,1,0)(1,1,1)24.}
\end{tablenotes}
\end{table}

%\listofchanges

\end{document}